\documentclass[useAMS,usenatbib,usegraphicx]{mn2e}
\title[Inward motions in massive star-forming cores]{A multi-transition molecular line study of inward motions towards massive star-forming cores}
\author[Yan Sun and Yu Gao]{Yan Sun$^{1, 2}$ and Yu Gao$^{1}$\thanks{E-mail: yugao@pmo.ac.cn}\\
$^{1}$Purple Mountain Observatory, Chinese Academy of Sciences, 2
West Beijing Road, Nanjing, Jiangsu 210008, China\\
$^{2}$Graduate School of the Chinese Academy of Sciences, Beijing 100080, China}

\begin{document}

\date{Accepted 2008 September 19. Received 2008 September 19; in original form 2008 April 25}

\pagerange{\pageref{firstpage}--\pageref{lastpage}} \pubyear{2008}

\maketitle

\label{firstpage}

\begin{abstract}
  A multi-transition 3$\,$mm molecular line single-pointing and mapping survey was carried out towards 29 massive star-forming cores in order to search for the signature of inward motions. Up to seven different transitions, optically thick lines HCO$^+$(1-0), CS(2-1), HNC(1-0), HCN(1-0), $^{12}$CO(1-0) and optically thin lines C$^{18}$O(1-0), $^{13}$CO(1-0) were observed towards each source. The normalized velocity differences ($\delta$V$_{\rm CS}$, $\delta$V$_{\rm HCO^{+}}$) between the peak velocities of optically thick lines and optically thin line C$^{18}$O(1-0) for each source were derived. Prominent inward motions are probably present in either HCO$^+$(1-0) or CS(2-1) or HNC(1-0) observations in most sources. Our observations show that there is a significant difference in the incidence of blue shifted line asymmetric line profiles between CS(2-1) and HCO$^+$(1-0). The HCO$^+$(1-0) shows the highest occurrence of obvious asymmetric feature, perhaps owing to different optical depth between CS(2-1) and HCO$^+$(1-0). HCO$^+$(1-0) appears to be the best inward motion tracer. The mapping observations of multiple line transitions enable us to identify six strong infall candidates G123.07-6.31, W75(OH), S235N, CEP-A, W3(OH), NGC7538. The infall signature is extended up to a linear scale $>0.2 \:$pc.
\end{abstract}


\section{Introduction}

Massive stars, once formed, dominate the luminous, kinematic, and chemical output of stars and their feedback play a dominant role in the evolution of molecular clouds and any subsequent star formation therein (Krumholz \& Bonnell 2007). Despite their importance the formation of massive stars is currently an unresolved problem. The difficulties lie in their short evolution time scale and complex star-forming environment in star cluster, typically embedded in the dense cores. 
Two main theories for high-mass star formation are mergers of several low-mass stars (Bonnell et al. 1998), which requires a high stellar density; and accretion of materials from the circumstellar disks, which requires high accretion rate to resist radiation pressure. The latter is similar to that of low-mass star formation (Shu et al. 1987), which accompanied by outflow during the process of gravitational collapse. It appears that more observations favor the latter scenario for massive star formation (e.g. Zhang et al. 2001; Beuther et al. 2002; Patel et al. 2003; Moscadelli et al. 2005; Jiang et al. 2005; Fuller et al. 2005; Beltr\'an et al. 2006). Nevertheless, intermediate theories that involve both accretions and mergers are now also quite commonly accepted (e.g. Whitney 2005; Bonnell \& Bate 2006).

  While accretion often can not be observed directly, its presence can be 
inferred from the presence of large scale infall and outflow (e.g., Klaassen \& Wilson 2007).
 Therefore the infall motion is one of the important elements to help understand 
the theories of massive star formation. These motions can be studied by 
investigating the profiles of optically thick molecular lines, which 
show the blue asymmetric structure, named ``blue profile'', a combination 
of a double peak with a brighter blue peak or a skewed single blue peak in 
optically thick lines (Mardones et al. 1997). Before a definite claim of 
collapse signature from blue profile, we should rule out all other 
possibilities. For example, we know double peak could be caused by two 
velocity components in the  line of sight. To rule out this possibility 
we must observe optically thin line that peaks at the self-absorption 
dip of optically thick line (Gregersen et al. 2000; Wu \& Evans 2003). 
Surely blue profile may also be caused by rotation or outflow. 
However, infall motion is the only process that would produce consistently the blue profile. 
Outflow and rotation only produce a blue asymmetric line profile along a 
particular line of sight to a source. Outflows strongly affect the line wings in the optically thick
molecular line along the outflow axes. Most infall candidates both show infall signatures and young outflows, 
but outflow does not seem to be responsible for blue asymmetry alone (Klaassen \& Wilson 2007).  
Rotation might cause blue asymmetric
line profile at one side of the rotation axis, but at the same time the red asymmetry could appear at the other side of the rotation axis. 
Mapping observation allows us to locate the center of the inflow motion and to 
identify dense cores that are simultaneously showing evidence for inflow motions and outflows (Wu et al. 2007).

Blue profile was detected in B335, a low-mass star formation
region (Zhou et al. 1993, 1994), which is well matched by the
standard model of ``inside-out'' gravitational collapse (Shu
1977). Subsequently, more and more blue profiles were found via
large surveys towards both young stellar objects (Mardones et al.
1997; Gregersen et al. 2000) and starless cores (Lee et al. 1999,
2001, 2004; Sohn et al. 2007) in low-mass star formation regions.
If massive stars are formed by accretion/collapse in a similar way
as low-mass stars, blue profile of optically-thick lines should
also be found towards them. Actually, these inflow signatures have
been reported in many massive sources (Welch et al. 1988; Dickel
\& Auer 1994; Wolf-Chase \& Gregersen 1997; Zhang, Ho \& Ohashi
1998). Especially, several large surveys of inflow signatures in
optical-thick lines towards massive young objects have been
carried out, which give the important evidence of
inflow/accretion/collapse (Wu \& Evans 2003; Fuller et al. 2005;
Klaassen \& Wilson 2007; Wu et al. 2007). In these surveys, the
infall velocity is found to be between $0.1$ km$\:$s$^-$$^1$ and 1
km$\:$s$^-$$^1$ and the mass infall rate is estimated to be
between 2$\times$10$^-$$^4$\textit{M}$_\odot$/yr and
$10^{-3}$\textit{M}$_\odot$/yr (e.g. Fuller et al. 2005).
 
   In this paper, we identify and study the inflow motion using several optical-thick lines including HCO$^+$(1-0), CS(2-1), HNC(1-0), HCN(1-0), and $^{12}$CO(1-0). Our investigation of the profiles of optically thick molecular lines is aimed to understand two questions: (1) Is the inflow motion in massive star formation cores similar to that in low-mass star formation cores? (2) Which optical thick molecular line is the best tracer for the infall motions in massive star-forming cores? Actually, such mapping observations towards collapsing massive star-forming regions, yet are seldom, enable us to identify the most probably infall candidates. Therefore, our observational study of multiple lines in 29 high-mass star-forming regions can help better characterize the detailed properties of massive star formation in the dense cores, reveal more information about the collapsing signatures and provide some strong infall candidates.

 After details of our sample selection and the related observations in the next section, the main results of this paper are presented in $\S$3; $\S$3.1 gives the single pointing observation results; $\S$3.2 presents the mapping results and analyses towards the infall candidates. We discuss our observational results in $\S$4 and summarize in $\S$5. 

\section{Sample \& Observations}

As a whole, the sources were selected by applying
the following criteria: (1) they are associated with water masers (24 sources)
or massive young stellar objects (5 sources); (2) they were included in the
Spitzer Space Telescope programs before
2007; (3) source distance is less than 4$\:$kpc. All of the 29 sources in our sample are listed in Table~1 with 22 sources (sources above the blank line in Table~1) selected from Shirley et al. 2003. To enlarge the sample, we also selected 7 sources (sources below the blank line in Table~1) from Spitzer programs that have images of massive star-forming regions. Among the 7 sources, S88 and IRAS19410 are also listed in Shirley et al. (2003) with different names as
S88B and G59.78+0.06 respectively, but the source center positions are slightly different.

There is evidence that water masers are a signature of recent or
ongoing high-mass star formation (e.g. Tofani et al. 1995). The
first criterion of our sample selections can ensure that all sources in our sample are
associated with massive star-forming regions. With high sensitivity and high spatial resolution at 
infrared wavelengths, the Infrared Array Camera (IRAC; Fazio et al. 2004) and Multiband Imaging Photometer
for Spitzer (MIPS; Rieke et al. 2004) see through dust into the birthplaces of stars and provide new insight to directly reveal the distribution
 of star formation sites and their environments in which stars form. IRAC data are available to
all of the tragets and the MIPS data are also available for most of them.
Considering the limited resolving power of the telescopes and the size of the infall region, we made some efforts to exclude distant sources. The 
source distances in the resulting sample are in a range from 0.6 to 4.0 kpc. The positions and 
distances are mainly determined from two literature search (see Table 1 for 
references). Most of the sample include cores of ultra-compact H{\sc ii} 
(UC~H{\sc ii}) regions, the sample also include cores of compact H{\sc ii} (CH{\sc ii}) and H{\sc ii} regions.

\begin{table*}
\centering
\begin{minipage}{170mm}
\caption{Source positions and rms $T_{\rm A}^{*}$ noise levels in spectra (with channel width 50 KHZ).}
\scriptsize
\begin{tabular}{lcccccccccc}
\hline
\hline
               & RA.            & Decl.        & Distance &      &  centroid&  CS~2-1 & HCO$^+$~1-0 & C$^{18}$O~1-0 & Map Size & Map Size\\
source           & (2000.0)       & (2000.0)     &  (kpc)   & H\sc{ii}?  &(\arcmin) &  $\rmn{K}$      & $\rmn{K}$        & $\rmn{K}$ & CS~2-1(\arcmin) & HCO$^+$~1-0(\arcmin)\\
\hline
G123.07-6.31     &   00:36:47.51  &   +63:29:02.08   &  2.2 & H\sc{ii}  & (0,0)  &   0.038 & 0.068 & 0.053  & 3$\times$3   &  3$\times$3\\
$\bullet$W3(OH)  &   02:27:04.69  &   +61:52:25.54   &  1.95 & UCH\sc{ii}& (0,0)  &   0.083 & 0.060 & 0.038  & 7$\times$7   &  7$\times$7\\
S231             &   05:39:12.91  &   +35:45:54.05   &  2.3 & \ldots    & (0,0)  &   0.061 & 0.084 & 0.072  & 2$\times$2   &  $\ldots$\\
S235             &   05:40:53.32  &   +35:41:48.76   &  1.6 & H\sc{ii}  & (0,0)  &   0.058 & 0.087 & 0.093  & 2$\times$2   &  $\ldots$\\
S252A            &   06:08:35.41  &   +20:39:02.96   &  1.5 & H\sc{ii}  & (0,0)  &   0.038 & 0.081 & 0.037  & 2$\times$2   &  $\ldots$\\
$\bullet$S255    &   06:12:53.72  &   +17:59:22.03   &  1.3 & UCH\sc{ii}& (0,0)  &         & 0.115 & 0.066  &              &  7$\times$7\\
G19.61-023       &   18:27:38.84  &   -11:56:27.44   &  4.0 & CH\sc{ii} & (0,0)  &   0.039 & 0.034 & 0.072  & $\ldots$      &  $\ldots$\\
G20.08-0.13      &   18:28:09.88  &   -11:28:48.17   &  3.4 & UCH\sc{ii}& (0,0)  &   0.040 & 0.046 & 0.086  & $\ldots$     &  $\ldots$ \\
G24.49-0.04      &   18:35:05.34  &   -07:31:26.98   &  3.5 & \ldots    & (0,0)  &         &       & 0.106  & $\ldots$     &  $\ldots$\\
$\bullet$W44     &   18:53:18.50  &   +01:14:56.65   &  3.7 & CH\sc{ii} & (0,0)  &   0.092 & 0.104 & 0.075  & 7$\times$7   &  7$\times$7\\
S76E             &   18:56:10.43  &   +07:53:14.11   &  2.1 & H\sc{ii}  & (0,0)  &   0.040 & 0.061 & 0.092  &3.5$\times$3  &  $\ldots$\\
G35.58-0.03      &   18:56:22.57  &   +02:20:27.75   &  3.5 & UCH\sc{ii}& (0,0)  &   0.038 & 0.043 & 0.076  &2$\times$2    &  $\ldots$\\
G35.20-0.74      &   18:59:12.73  &   +01:40:40.75   &  3.3 & H\sc{ii}  & (0,0)  &   0.027 & 0.027 & 0.091  & $\ldots$     &  $\ldots$\\
OH43.80-0.13     &   19:11:54.27  &   +09:35:55.38   &  2.7 & UCH\sc{ii}& (0,0)  &   0.039 & 0.052 & 0.078  &2$\times$2    &  $\ldots$\\
$\bullet$ S87    &   19:46:20.45  &   +24:35:34.41   &  1.9 & UCH\sc{ii}& (0,0)  &   0.111 & 0.088 & 0.076  & 7$\times$7   &  7$\times$7\\
$\bullet$S106    &   20:27:25.74  &   +37:22:51.82   &  0.6 & UCH\sc{ii}& (0,0)  &   0.070 & 0.079 & 0.030  & 7$\times$7   &  7$\times$7\\
$\bullet$W75N    &   20:38:36.93  &   +42:37:37.52   &  3.0 & UCH\sc{ii}& (0,0)  &   0.091 & 0.071 & 0.047  & 7$\times$7   &  7$\times$7\\
DR21S            &   20:39:00.80  &   +42:19:29.84   &  3.0 & UCH\sc{ii}& (0,0)  &   0.038 & 0.076 & 0.137  & 2$\times$2   &  2$\times$2\\
$\bullet$W75(OH) &   20:39:01.01  &   +42:22:49.84   &  3.0 & \ldots    & (0,0)  &   0.118 & 0.090 & 0.085  & 7$\times$7   &  7$\times$7\\
$\bullet$BFS11-B &   21:43:06.68  &   +66:07:04.11   &  2.0 & \ldots    & (0,0)  &   0.085 & 0.107 & 0.065  & 7$\times$7   &  7$\times$7\\
$\bullet$Cep-A   &   22:56:18.14  &   +62:01:46.35   &  0.7 & UCH\sc{ii}& (0,0)  &   0.117 & 0.089 & 0.024  & 7$\times$7   &  7$\times$7\\
$\bullet$NGC7538 &   23:13:44.86  &   +61:26:50.71   &  2.8 & UCH\sc{ii}& (0,0)  &   0.232 & 0.073 & 0.032  & 7$\times$7   &  7$\times$7\\
                 &                &                  &      &           &        &         &       &        &              &\\
AFGL4029         &   03:02:00.00  &   +60:33:49.00   &  2.2 & UCH\sc{ii}& (-3,-4) &   0.054 & 0.061 & 0.063  &2.5$\times$2.5& 2.5$\times$3\\
AFGL5142         &   05:30:48.00  &   +33:47:53.80   &  1.8 & H\sc{ii} & (0,0)  &   0.041 & 0.073 & 0.040  &2$\times$2    & 2$\times$2\\
S235N            &   05:41:00.00  &   +35:48:04.00   &  1.6 &\ldots & (6,2)  &   0.042 & 0.068 & 0.053  &2$\times$2    & 2$\times$2 \\
G192             &   05:58:17.07  &   +16:31:58.00   &  2.0 & UCH\sc{ii}& (-1,0)  &   0.025 & 0.052 & 0.070  &2$\times$2    & $\ldots$\\
IRAS19410        &   19:43:11.00  &   +23:44:06.00   &  2.2 & UCH\sc{ii}& (0,0)  &   0.031 & 0.040 & 0.028  &3$\times$3    &2$\times$2\\
S88              &   19:46:43.00  &   +25:12:14.00   &  2.1 & UCH\sc{ii}& (1,1)  &   0.047 & 0.072 & 0.059  &2$\times$2    &2$\times$2\\
IRAS20126        &   20:14:25.80  &   +41:13:33.00   &  1.7 & \ldots& (0,0)  &   0.039 & 0.114 & 0.094  &3.5$\times$3  &$\ldots$ \\

 \hline
\end{tabular}
\medskip

NOTE.-positions and distances of sources above the blank line are from Shirley et al. 2003, except for W3(OH) and S106. The distance of W3(OH) is from Xu et al. 2006. The distance of S106 is from Eiroa Elasser \& Lahulla 1979; Ridge et al. 2003. For the last eight sources, positions are from Spitzer observing center, distances are from Carral et al. 1999, except for IRAS19410. The distance of IRAS19410 is from Xu et al. 2007. Note that S235 has a relatively extended distribution, the north region was indicated by S235N. Centroid is the $^{13}$CO(1-0) peak position of each source. Columns 7,8,9 are rms  $T_{\rm A}^{*}$ noise levels twords $^{13}$CO(1-0) peak position of each source in CS(2-1), HCO$^+$(1-0), C$^{18}$O(1-0) respectively. Symbol $\ldots$ in the columns 10 and 11 denoted only single point was observed.
\end{minipage}
\end{table*}
\normalsize

Our observations were carried out using the 14$\:$m telescope of
Purple Mountain Observatory (PMO) in Delingha, China. The
position-switch mode was used, the pointing accuracy is estimated
to be better than 9\arcsec. One SIS receivers operating at
$\sim$110 GHz were used in the observations along with an
acoustic-optic spectrometer (AOS) back end with 1024 channels. The
standard chopper wheel calibration was used during the
observations to get the antenna temperature, $T_{\rm A}^{*}$, which
has been corrected for atmospheric absorption. The typical system
temperatures $T_{\rm sys}$ is about 250-350$\:$K. At central source
position, the accumulated integration time towards each source is
larger than 10 minutes and the signal-to-noise ratio is larger than
10. We carried out our mapping observation only towards strong
emission ($T_{\rm A}^{*}>0.5\:K$ in HCO$^+$(1-0) and CS(2-1)) sources
with line asymmetry for the sake of the observational efficiency.
The positions of CO emission peak (which were marked in Table 1)
have the highest priority in further HCO$^+$(1-0) and CS(2-1)
observations. 

The mapping step is 30$\arcsec$ for CS(2-1), HCO$^+$(1-0),
and 1$\arcmin$ ($\sim$ beam size, for our source distance, corresponding to 
linear resolution between 0.16$\,$pc and 1.07$\,$pc) for $^{12}$CO(1-0), $^{13}$CO(1-0), C$^{18}$O(1-0). To
complement our study, we have also used some data from Ma, Gao, \&
Wu (2008 in prep.) including CS(2-1), HCO$^+$(1-0), HNC(1-0) and
HCN(1-0) mapping data. These ten sources are marked with $\bullet$ in
Table~1 and they were observed with the SEQUOIA on the former Five
Colleges Radio Astronomy Observatory (FCRAO) 14m telescope with a
25$\arcsec$ on-the-fly mapping step. 
The detailed observation log is listed in Table~2.
\begin{table*}
\centering
\begin{minipage}{140mm}
\caption{Observed lines and telescope parameters.}
\scriptsize
\begin{tabular}{lccrcccc}
\hline
\hline
Line     &  Data   &   Frequency   &  Telescope & FWHM
& $\eta$$_m$$_b$\footnote{$\eta$$_m$$_b$ is the main beam efficiency.} & $\Delta$v & T$_{sys}$($T_{\rm A}^{*}$)\\
   &  (UT) & (GHZ) &  & (arcsec) &  & (km$\:$s$^-$$^1$) & K \\
   \hline
$^{12}$CO(1-0)  & 2005 Nov & 115.271204  & PMO  14m & 55 & 0.72 &0.37  &  \\
                & 2006 Nov & 115.271204  & PMO  14m & 55 & 0.57 &0.37  &  \\
$^{13}$CO(1-0)  & 2005 Nov & 110.201353  & PMO  14m & 55 & 0.72 &0.11  &  \\
                & 2006 Nov & 110.201353  & PMO  14m & 55 & 0.57 &0.11  &  \raisebox{1.3mm}[0pt]{200-350}\\
$^1$$^8$CO(1-0) & 2005 Nov & 109.782173  & PMO   14m & 55 &0.72&0.12   &  \\
                & 2006 Nov & 109.782173  & PMO   14m & 55 &0.57&0.12   &  \\
HCO$^+$(1-0)    & 2007 Jun &  89.1885260 & PMO   14m & 55&0.62&0.14    &  280-400\\
                & 2005 May &  89.1885180 & FCRAO 14m & 57&0.53&0.16    &  160-210\\
CS(2-1)         & 2007 May & 97.9809530  & PMO   14m & 55 & 0.62 &0.13 &  160-250\\
                & 2005 May & 97.9810070  & FCRAO 14m & 54 & 0.49 &0.15 &  120-210 \\
HNC(1-0)        & 2005 May &90.6635720   &FCRAO  14m &57 &0.53 &0.16   &  160-210\\
HCN(1-0)        & 2005 May &88.6318473   &FCRAO  14m &57 &0.53& 0.17   &  120-200\\
\hline
\end{tabular}
\end{minipage}
\end{table*}
\normalsize

   The spectral data were reduced and analyzed with CLASS and IDL software. Linear baselines were removed from all spectra. The rms levels for each spectrum at the CO emission peak are also listed in Table~1. The complete description of our multiple molecular line observations and all data presentation will be given in a forthcoming paper (Sun \& Gao 2008 in prep.)

There is little overlap in the samples of previous observational search for infall signatures, particularly the mapping studies. 
Two major surveys of inflow motion observations that have
some source overlaps with our sample are  Wu \& Evans (2003) and Klaassen \& Wilson (2007).  There are fifteen sources that overlap between 
our HCO$^+$(1-0) mapping observations and that of single-pointing HCN(3-2) observation of Wu \& Evans (2003). Yet, only five sources in our sample overlap with the single-pointing  HCO$^+$(4-3) observation of Klaassen \& Wilson (2007).

\section{Analysis \& Results }

 All targets are detected in HCO$^+$(1-0), CS(2-1), C$^{18}$O(1-0), $^{12}$CO(1-0), $^{13}$CO(1-0), except for G24.49-0.04 owing to a wrong frequency setup (a wrong V$_{\rm LSR}$ was taken). Most sources are pretty strong with $T_{A}^{*}$ of HCO$^+$(1-0), CS(2-1) mainly in a range of 1-4.5$\:$K. But several sources (G19.61-023, G20.08-0.13, G35.58-0.03, G35.20-0.74) show fairly weak emission ($T_{A}^{*}<0.3\:$K) in HCO$^+$(1-0), and one source (G35.20-0.74) shows quite weak emission ($T_{\rm A}^{*}<0.2\:$K) in CS(2-1). Among the 29 dense cores observed, S255 shows a very marginal detection in CS emission and S76E has two velocity components in both C$^{18}$O(1-0) and $^{13}$CO(1-0) and is difficult to determine the systematic velocity. Therefore, these sources (G24.49-0.04, S255, S76E) are excluded from the following analysis. We only give the results derived from CS(2-1) and HCO$^+$(1-0) as optically thick spectral lines towards 26 sources below since only ten sources have both HCN(1-0) and HNC(1-0) observations.

\subsection{Blue profile identification}

  To quantify the blue asymmetry of a line, we use an asymmetry parameter $\delta$V (pioneered by Mardones et al. 1997) defined as the difference between the peak velocities of an optically thick line V(thick) and an optically thin line V(thin) in units of the optically thin line FWHM (Full Width at Half Maximum) dV(thin): $\delta V={V(thick)-V(thin)\over dV(thin)}$. Mardones et al. (1997) adopted a criterion $\delta V<-$0.25 to indicate blue asymmetry and $\delta V>0.25$ to indicate red asymmetry. They chose the threshold value 0.25 at about 5 times the typical rms error in $\delta$V in order to exclude the random contributions (Mardones et al. 1997). We use the same criterion for consistency with their analysis. Because of the asymmetric profiles for many of optical-thick lines, they cannot be fitted by Gaussian function, the peaks of the line profiles of CS(2-1), HCO$^+$(1-0), and HNC(1-0) are identified with the highest flux by cursor. The uncertainties of the velocities and the antenna temperatures at these peaks were estimated by several measurements and the baseline fitting noises. The optically thin line (C$^{18}$O) shows an ideal Gaussian profile and is fitted by the Gaussian function to obtain the peak velocity and FWHM. All these parameters towards the $^{13}$CO(1-0) peak position are listed in Table~3 which includes the peak velocities of CS(2-1), HCO$^+(1-0)$, C$^{18}$O(1-0), the FWHM of C$^{18}$O(1-0), the calculated  asymmetry parameter $\delta$V of CS(2-1) and HCO$^+$(1-0), and the uncertainties of these parameters (in the parentheses) for all sources.

 The blue profile caused by infall motion also requires $T_{\rm A}^{*}(B)/T_{\rm A}^{*}(R)>1+\sigma$ (i.e., the blue peak is stronger than the red peak, hereafter T$_{\rm B}/T_{\rm R}\equiv T_{\rm A}^{*}(B)/T_{\rm A}^{*}(R)$). These ratios of CS(2-1) and HCO$^+(1-0)$ towards the $^{13}$CO(1-0) peak position are available in Table~3. The peak antenna temperatures of CS(2-1) and HCO$^+$(1-0) towards each source are also listed in table 3 columns 8 and 10, respectively. Based on the methods that we used to derive the ratio of $T_{\rm B}/T_{\rm R}$, the sources can be classified into three groups: (1) sources show obvious two peaks, we can accurately measure the ratio of  $T_{\rm B}/T_{\rm R}$; (2) sources with asymmetry profiles but do not show obviously two peaks (both blue and red), thus only one peak (either blue or red velocity offset) can be identified. We simply chose the same velocity offset to the systematic velocity as the other peak, so $T_{\rm B}/T_{\rm R}$ can not be exactly derived. We mark ``$\sim$'' before such sources; (3) sources almost don't show any asymmetry, neither the red nor blue component can be identified, so the intensity ratio $T_{\rm B}/T_{\rm R}$ was flagged with ``$\ldots$''.

In our sample, four blue profiles were identified in CS(2-1),
fourteen blue profiles were identified by HCO$^+$(1-0), three sources
show blue profiles both in CS(2-1) and HCO$^+$(1-0) spectra.
Within the error bars, the parameters ($\delta$V,
$T_{\rm B}/T_{\rm R}$) of several sources (W44, G192,
G35.20-0.74,S88) are approaching the criterion, so they are possible
candidates of the blue profiles or red profiles and are marked with ``?'' in Table
3. Note that S87 and G35.20-0.74 show different asymmetry between
CS(2-1) and HCO$^+$(1-0). S87 also shows different asymmetry at
different positions. At offsets (0$\arcsec$, 25$\arcsec$), (0$\arcsec$, -25$\arcsec$), (25$\arcsec$, 0$\arcsec$), (-25$\arcsec$, 0$\arcsec$), it shows red asymmetry, blue asymmetry, no asymmetry, blue asymmetry in HCO$^+$(1-0) and red asymmetry, no asymmetry, red asymmetry, red asymmetry in CS(2-1), respectively.  These might suggest a complicated kinematics in the
mapped regions in S87 (c.f. Xue \& Wu 2008, where they claimed that the asymmetric line profiles in S87 should result from two clouds at slightly different velocities). Figure 1 presents the statistics of the asymmetry
parameters measured towards the $^{13}$CO(1-0) peak position of the 26 sources observed in both
HCO$^+$(1-0) and CS(2-1) mentioned before. It's clear that the blue profile
predominance was shown, particularly in HCO$^+$(1-0). Because infall is the only process 
that would produce consistently the blue profile, its predominance indicates that 
the infall motions are really present in massive star-forming regions. A comparison of
the asymmetry parameter $\delta$V in CS(2-1) and HCO$^+$(1-0) is
shown in Figure~2, which shows weak correlation between $\delta$v in the two
molecular lines. As noted above, G35.20-0.74 and S87
show different asymmetry between CS(2-1) and HCO$^+$(1-0) and also
greatly affect the correlation between $\delta$V of the two
lines. Generally, there are more sources with blue asymmetric
profiles in HCO$^+$(1-0) line than those in CS(2-1). We will
further discuss the difference between the two lines in $\S$4.

%The quantity ``excess'' is defined by Mardones et al. (1997) as $E=(N_B-N_R)/N_T$, where $N_{B}$ and $N_{R}$ are the numbers of blue and red profiles in the total number of sources $N_{T}$. In our sample, the excess in our sample is E=0.07 and 0.54 for CS(2-1) and HCO$^+$(1-0) respectively. 
\begin{figure}
%%\vspace{2mm}
\centering
\includegraphics[scale=0.25]{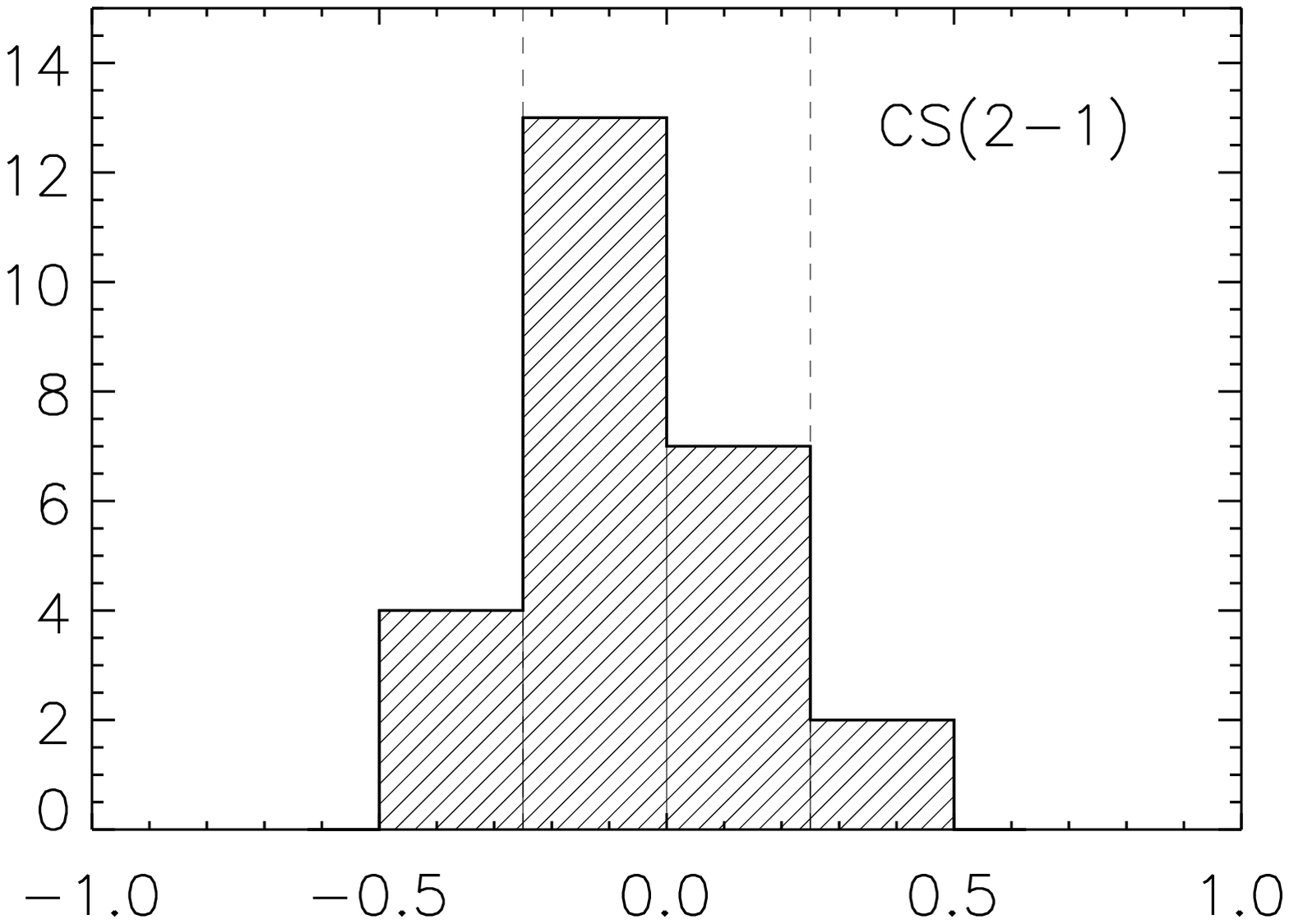}%
\includegraphics[scale=0.25]{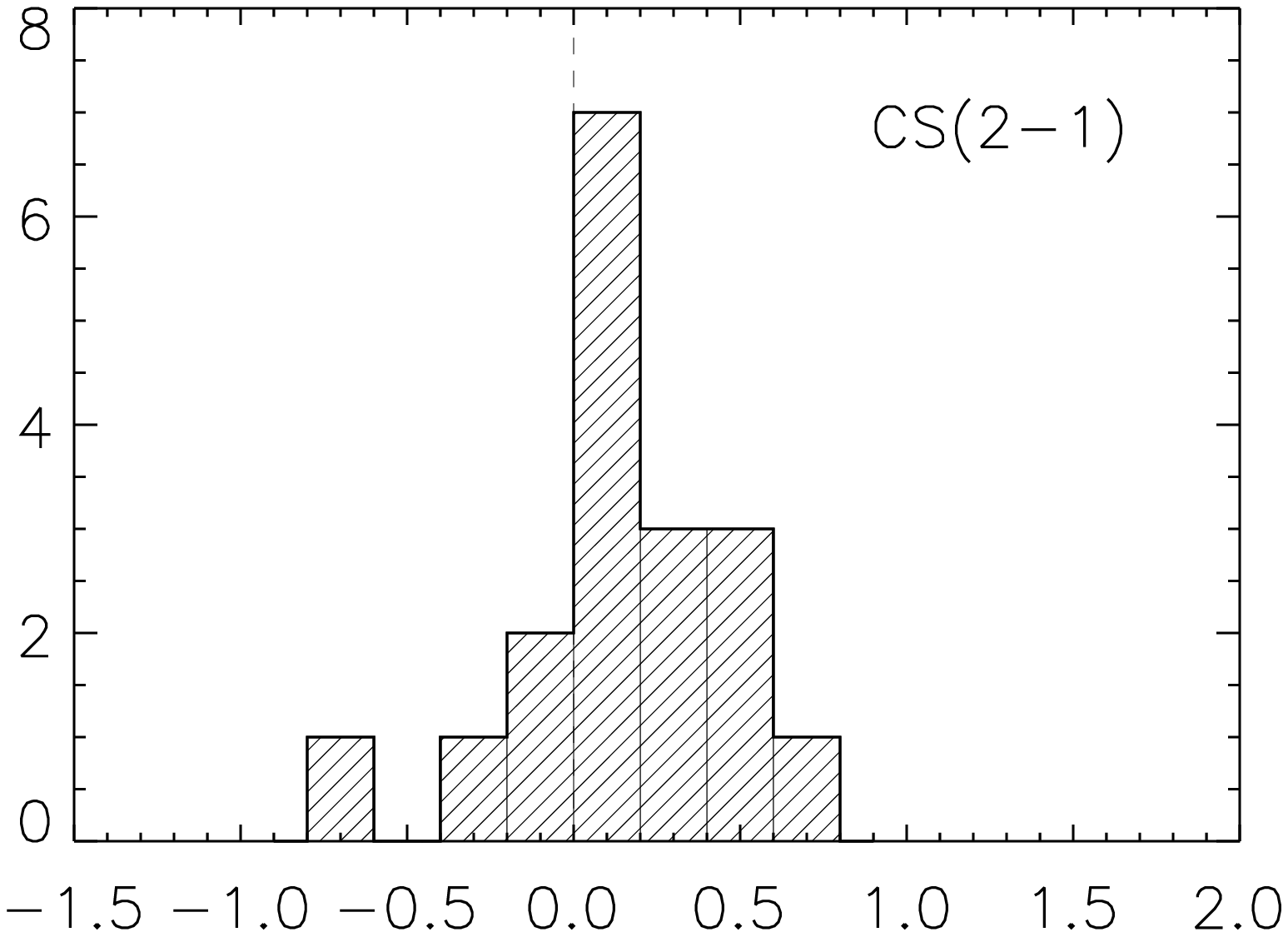}
\includegraphics[scale=0.25]{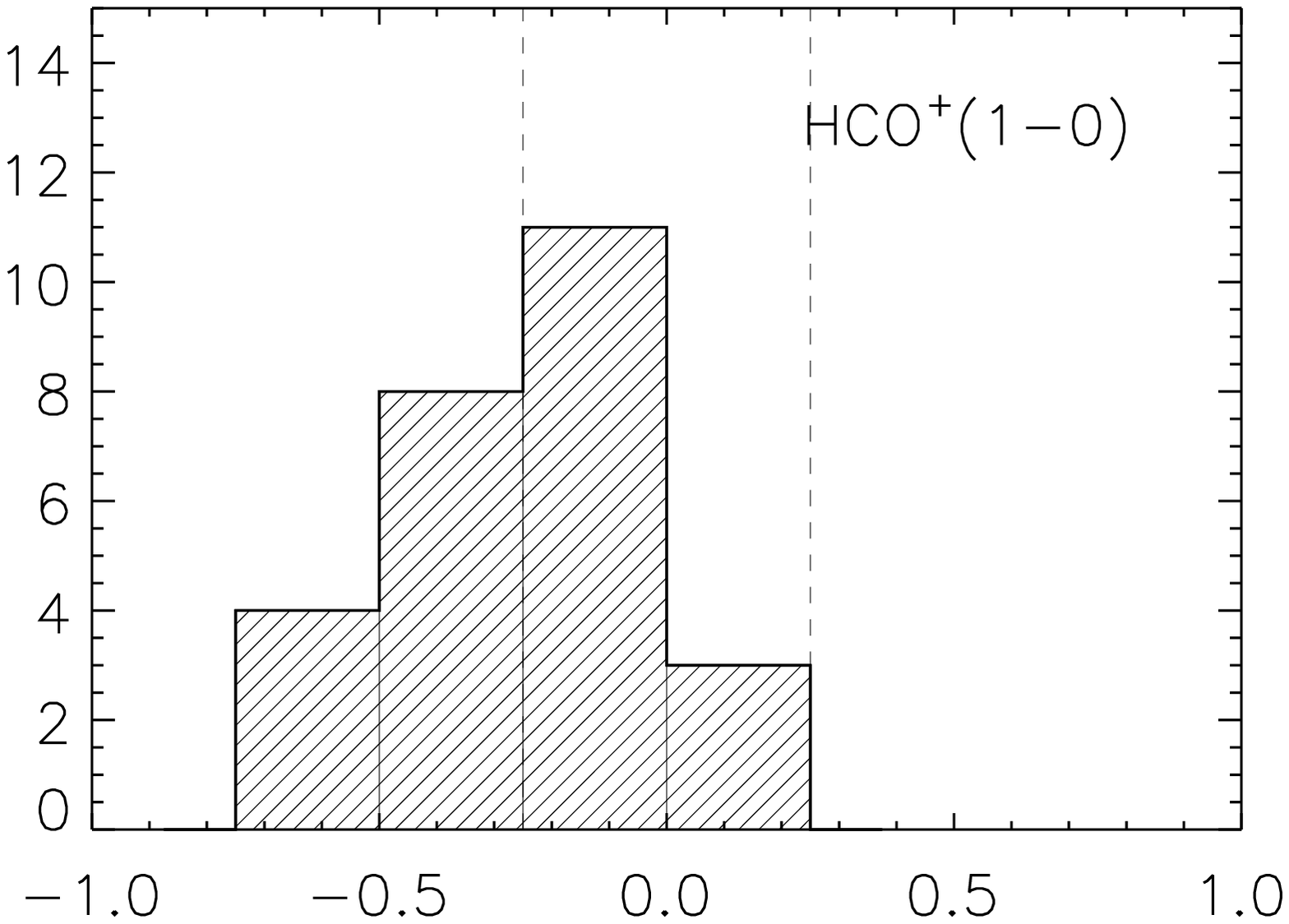}%
\includegraphics[scale=0.25]{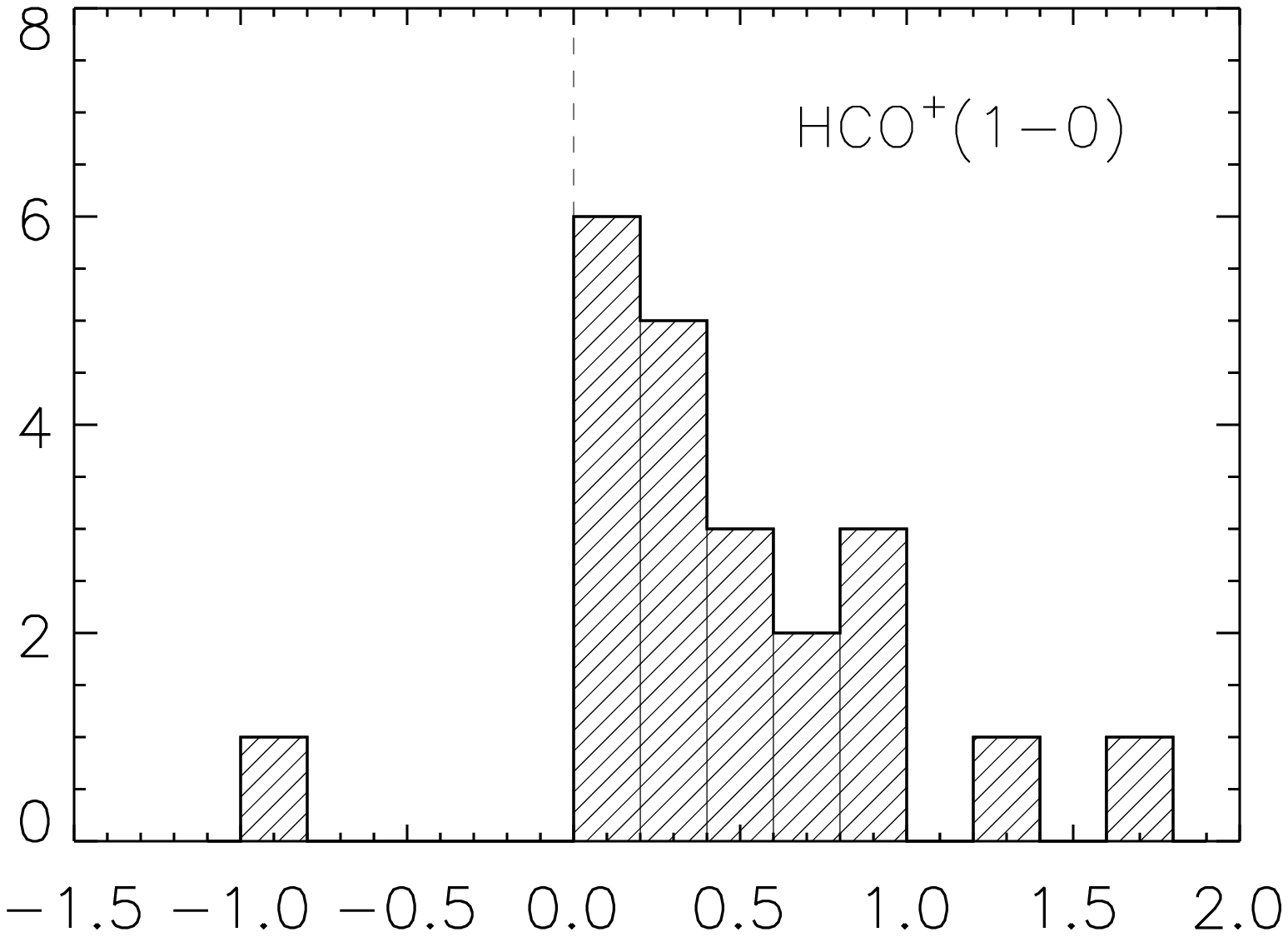}
%\vspace{1mm}
\caption{Distribution of $\delta$V (left, the line asymmetry parameter) and 
log[$T_{\rm A}^{*}$(B)/$T_{\rm A}^{*}$(R)] (right, ratio of blue to red peak intensity). The top and bottom panels show results
from CS(2-1) and HCO$^+$(1-0) respectively.}
\end{figure}

\begin{table*}
\centering
\begin{minipage}[t]{180mm}
\caption{The Derived line parameters and profiles of the observed
sources. Quantities in parentheses give the uncertainties in units
of 0.01.}
\scriptsize
\begin{tabular}{lccccccccccc}
 \hline
 \hline
source \footnote{Sources marked with $\diamond$ and  * are considered to be probably and strong  infall candidates, respectively.}     &  $V_{thick}$ &  $V_{thick}$ &   $V_{thin}$  &   $\Delta$V   &  $\delta$v &  $\delta$v & T$_{peak}$ & $T_{B}$/$T_{R}$ & T$_{peak}$ & $T_{B}$/$T_{R}$& profile \footnote{Profile is judged from our CS(2-1), HCO$^+$(1-0) observations and HCN(3-2) observation of Wu \& Evans (2003), respectively. B denotes blue profile, R denotes red profile, N denotes neither blue nor red, $\sim$ denotes no HCN(3-2) data} \\
& CS~2-1 & HCO$^+$~1-0 & C$^{18}$O~1-0 & C$^{18}$O~1-0 & CS~2-1 & HCO$^+$~1-0 & CS~2-1 & CS~2-1 & HCO$^+$~1-0 & HCO$^+$~1-0 & \\
 &      km$\:$$s^{-1}$ & km$\:$$s^{-1}$ & km$\:$$s^{-1}$ & km$\:$$s^{-1}$ & & & K &  & K & \\
\hline
$*$g123.07-6.31   & -18.04(04)  & -18.35(01)   & -17.55(02)   & 1.80(04)  & -0.27(03)  &  -0.44(02) & 1.52 & 1.54       & 2.95   &  1.79      &  B,B,B \\
$*$W3(OH)         & -47.53(06)  & -48.76(07)   & -47.15(02)   & 4.16(05)  & -0.09(02)  &  -0.39(02) & 3.73 & $\sim$1.08 & 2.87   &  2.34      &  N,B,B \\
S231              & -17.14(03)  & -17.33(01)   & -16.76(03)   & 3.66(07)  & -0.10(01)  &  -0.16(01) & 1.31 & $\sim$1.08 & 2.18   & $\sim$1.09 &  N,N,R \\
S235              & -16.83(09)  & -17.30(04)   & -16.85(04)   & 2.38(09)  &  0.01(04)  &  -0.19(02) & 2.12 & $\ldots$   & 2.52   & $\sim$1.47 &  N,N,N \\
S252A             &   8.95(05)  &   8.60(10)   &   8.53(01)   & 2.44(03)  &  0.17(02)  &   0.03(04) & 1.76 & $\sim$0.91 & 1.56   & $\ldots$   &  N,N,R \\
G19.61-0.23       &  42.79(12)  &  40.31(06)   &  42.75(12)   & 6.61(34)  &  0.01(03)  &  -0.37(03) & 0.62 & $\ldots$   & 0.32   & $\sim$1.58 &  N,B,N \\
G20.08-0.13       &  41.79(14)  &  41.29(07)   &  42.13(13)   & 4.83(39)  & -0.07(04)  &  -0.17(03) & 0.53 & $\ldots$   & 0.29   & $\sim$1.53 &  N,N,$\sim$ \\
$\diamond$W44     &  58.02(13)  &  56.51(07)   &  57.76(02)   & 5.20(04)  &  0.05(03)  &  -0.24(01) & 3.27 & $\ldots$   & 2.96   & $\sim$4.96 &  N,B?B \\
G35.58-0.03       &  53.29(04)  &  54.00(15)   &  53.01(02)   & 6.16(06)  &  0.05(01)  &   0.13(01) & 1.01 & $\sim$0.94 & 0.34   & $\sim$0.39 &  N,N,N \\
G35.20-0.74       &  37.50(04)  &  36.25(09)   &  37.09(05)   & 1.35(09)  &  0.30(05)  &  -0.62(09) & 0.16 & $\sim$0.79 & 0.17   & $\sim$1.05 &  R?B?$\sim$ \\
OH43.80-0.13      &  43.88(06)  &  43.82(12)   &  44.41(12)   & 5.30(21)  & -0.10(03)  &  -0.11(03) & 0.72 & $\sim$1.19 & 0.80   & $\sim$1.20 &  N,N,$\sim$ \\
S87               &  23.35(06)  &  21.20(07)   &  22.42(09)   & 3.39(16)  &  0.27(03)  &  -0.36(04) & 2.15 & 0.47       & 3.28   &  1.26      &  R,B,$\sim$ \\
$\diamond$S106    &  -1.47(12)  &  -1.85(07)   &  -1.01(01)   & 2.47(03)  & -0.19(05)  &  -0.34(03) & 1.74 & $\sim$1.58 & 1.63   & $\sim$1.33 &  N,B,$\sim$ \\
W75N              &   9.25(07)  &   9.55(05)   &   9.69(02)   & 3.47(05)  & -0.13(02)  &  -0.04(02) & 3.30 & $\sim$1.00 & 2.97   & $\ldots$   &  N,N,R \\
DR21S             &  -2.59(03)  &  -4.47(05)   &  -2.78(04)   & 3.12(09)  &  0.06(02)  &  -0.54(03) & 2.88 & $\ldots$   & 2.41   &  1.35      &  N,B,B \\
$*$W75(OH)        &  -4.28(07)  &  -5.03(07)   &  -3.06(03)   & 2.97(07)  & -0.41(03)  &  -0.66(03) & 4.36 & 1.88       & 4.11   &  2.00      &  B,B,B \\
BFS11-B           & -10.05(05)  & -10.25(05)   & -10.22(02)   & 1.58(08)  &  0.11(03)  &  -0.02(03) & 1.50 & $\sim$1.25 & 2.35   &  $\ldots$  &  N,N,$\sim$ \\
$*$Cep-A          & -11.44(16)  & -12.71(10)   & -10.62(01)   & 3.56(03)  & -0.23(05)  &  -0.59(03) & 2.43 & 1.19       & 2.29   &  2.29      &  N,B,B \\
$*$NGC7538        & -57.36(14)  & -57.81(21)   & -56.00(03)   & 5.40(07)  & -0.25(03)  &  -0.34(04) & 5.16 & $\sim$1.48 & 3.96   & $\sim$3.42 &  B,B,$\sim$ \\
AFGL4029          & -38.35(01)  & -38.49(03)   & -38.08(04)   & 2.19(10)  & -0.12(02)  &  -0.19(02) & 1.89 & $\sim$1.19 & 1.82   & $\sim$1.22 &  N,N,$\sim$ \\
AFGL5142          &  -3.51(02)  &  -3.84(02)   &  -3.37(02)   & 2.75(04)  & -0.05(01)  &  -0.17(01) & 3.67 & $\ldots$   & 3.71   & $\sim$1.31 &  N,N,$\sim$ \\
$*$S235N          & -20.23(10)  & -20.83(04)   & -19.62(05)   & 2.83(11)  & -0.22(04)  &  -0.43(03) & 1.08 & $\sim$1.28 & 1.98   &  2.68      &  N,B,$\sim$ \\
G192              &   5.61(06)  &   6.28(05)   &   6.18(05)   & 2.10(10)  & -0.27(04)  &   0.05(03) & 0.31 & $\sim$1.63 & 0.70   & 1.09       &  B?N,$\sim$ \\
IRAS19410         &  22.54(01)  &  21.50(04)   &  22.57(01)   & 2.35(02)  & -0.01(01)  &  -0.46(02) & 1.82 & $\ldots$   & 1.50   & 1.06       &  N,B,B \\
S88               &  21.90(06)  &  21.38(11)   &  22.04(03)   & 2.69(06)  & -0.05(02)  &  -0.25(04) & 0.91 & $\ldots$   & 2.43   &  2.18      &  N,B?R \\
IRAS20126         &  -3.88(03)  &  -3.65(05)   &  -3.58(06)   & 2.45(16)  & -0.12(03)  &  -0.03(03) & 1.06 & $\sim$1.16 & 2.24   & $\ldots$   &  N,N,$\sim$ \\
\hline
\end{tabular}\\
\end{minipage}
\end{table*}
\normalsize

\begin{figure}
%%\vspace{2mm}
\centering
\includegraphics[scale=0.5]{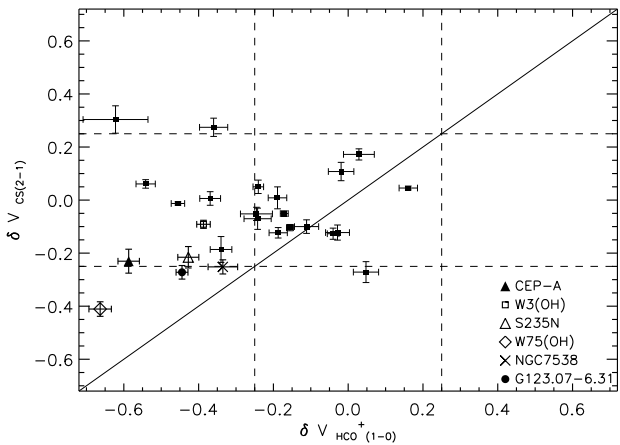}
\caption{Comparison of the measured asymmetry in HCO$^+$(1-0) and
CS(2-1). The dashed lines mark $|\delta V|=0.25$ and the line of
equality is shown in solid.}
\end{figure}

\subsection{Infall candidates}

The mapping observations of multiple transitions could offer
detailed investigation to provide strong evidence on whether the
core is an infall candidate. It has been widely accepted that the
infall candidates must show obvious blue profile at least in one
optically thick line, no red profile in other infall tracers and
no spacial difference in the mapping observation. Based on all of
these, we have found three probable infall candidates (S106, W44,
IRAS19410) and six strong infall candidates (G123.07-6.31,
W75(OH), S235N, CEP-A, W3(OH), NGC7538). The strong infall candidates
are marked in Table~3 and Figure~2.

Figures~3-8~show the HCO$^+$(1-0) mapping observations towards the six most strong infall 
candidates (Figs.4-7 are presented with only central parts of the HCO$^+$(1-0) maps). Most of the data show obvious asymmetric profiles 
and all of spectral lines observed towards the $^{13}$CO(1-0) emission peak position. 
The six infall candidates show consistent blue profile in at least one optically thick spectral line. 
Most of them show high velocity wing emissions both in HCO$^+$(1-0) and $^{12}$CO(1-0). These might suggest that besides infall motions outflows are also 
present in the infall candidates.  
Klaassen \& Wilson (2007) also found that seven out of their nine infall candidates show SiO detections. 
All these confirm that inflow was often accompanied by outflow in the process of massive star forming process.
In these strong infall candidates a large scale blue asymmetry was present. 
From our maps we estimated the extent of the infall signature to be up to 2$\arcmin\times2\arcmin$  (at least
0.66~pc for G123.07-6.31, 0.59~pc for W3(OH), $1.03\:$pc for W75(OH), $0.21\:$pc for CEP-A, 0.84~pc for NGC7538, and 0.34~pc for S235N).

We here choose G123.07-6.31 as a typical example of
infall candidate to illustrate our observations (Figure 3). In Figure~3a, 
a signature of extended inward motions were detected in all the 
mapping region of our HCO$^+$(1-0) observation, detailed analysis
was needed to better understand the spatial variation of the
infall speed of the core. Figure 3b shows all of molecular lines
observed towards the maser site (0,0), and outer region of the
core (1$\arcmin$,-1$\arcmin$). Obvious blue asymmetry is presented in both
CS(2-1) and HCO$^+$(1-0). Along $\Delta$decl. =0 the
position-velocity (P-V) diagram of HCO$^+$(1-0) shows strong blue
asymmetry (Figure~3c top) while the P-V diagram of C$^{18}$O(1-0)
doesn't show any asymmetry (Figure~3c bottom).

\begin{figure}
%%\vspace{2mm}
\centering
\begin{minipage}[c]{0.5\textwidth}
\centering
\includegraphics[width=7.9cm]{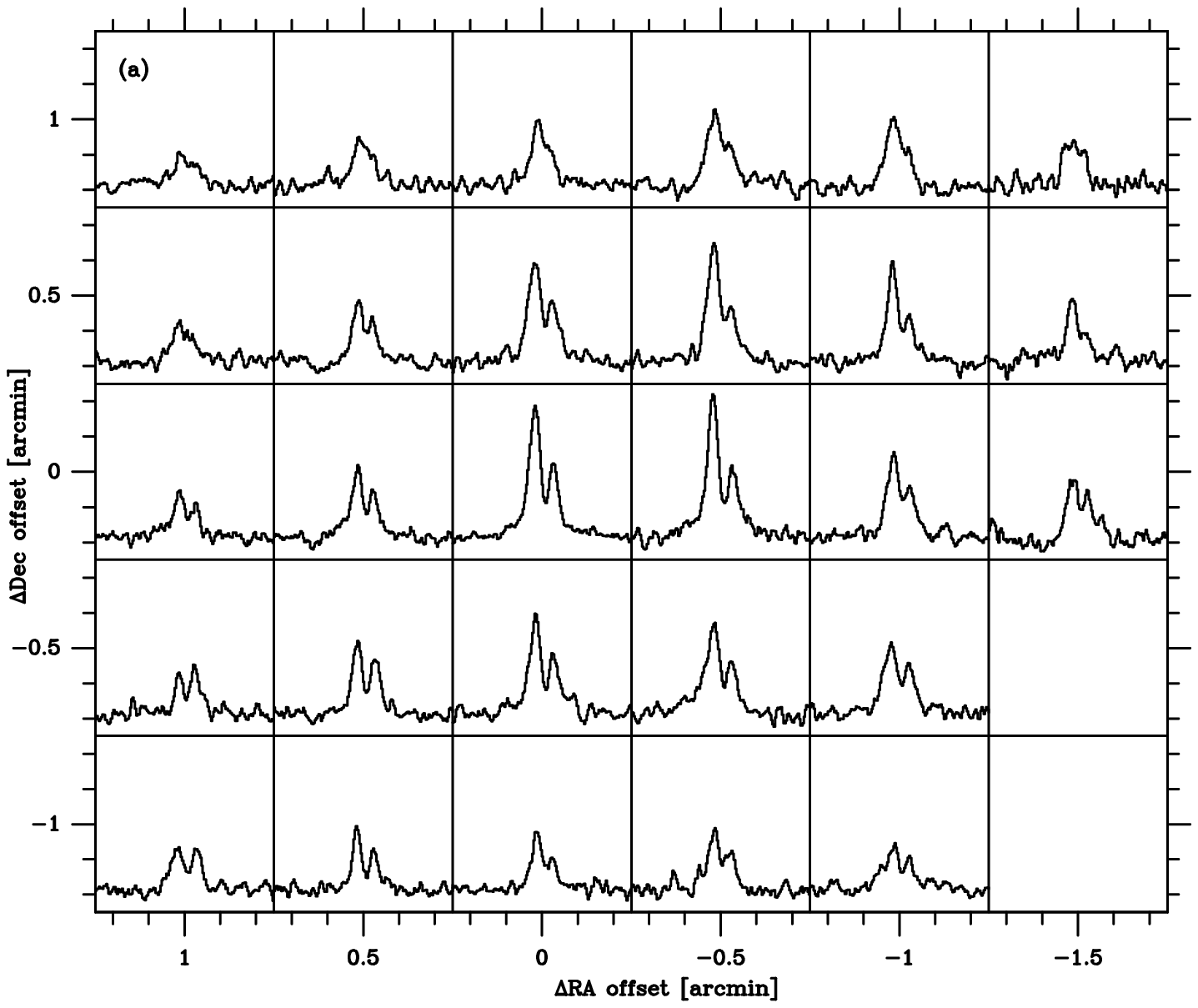}
\end{minipage}\\[0pt]
\begin{minipage}[c]{0.5\textwidth}
\centering
\includegraphics[width=7.9cm]{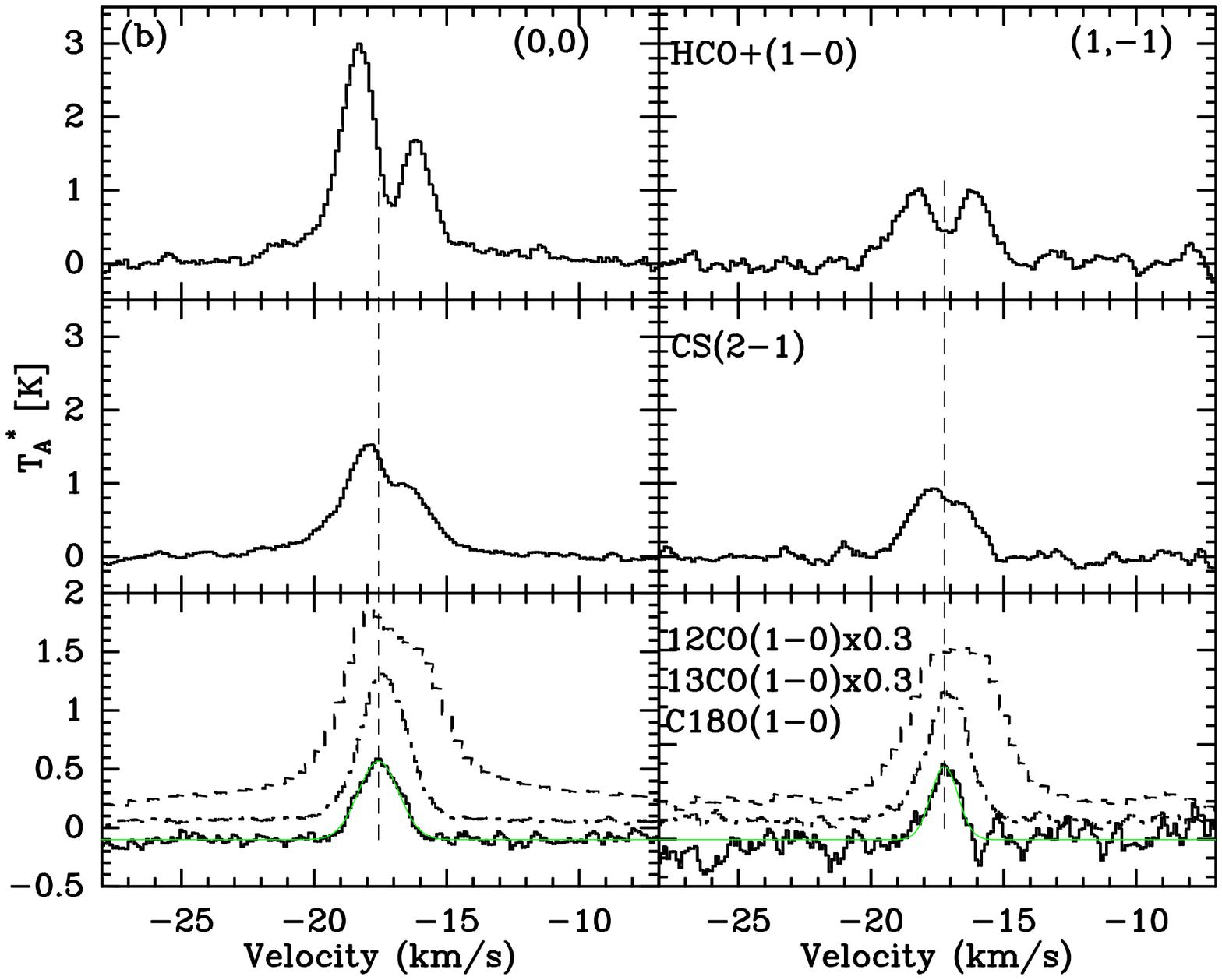}
\end{minipage}\\[0pt]
\begin{minipage}[c]{0.5\textwidth}
\centering
\includegraphics[width=7.9cm]{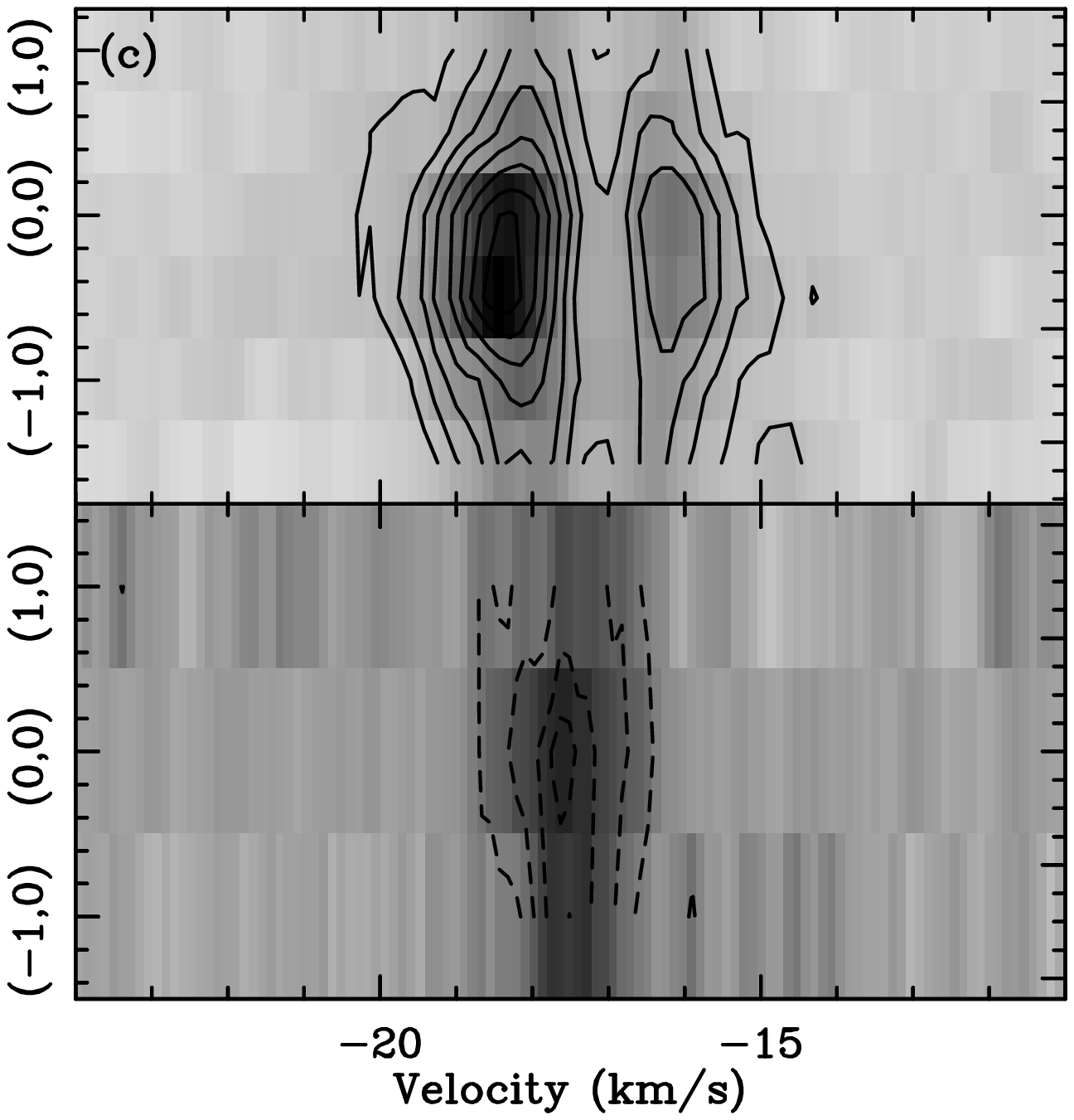}
\end{minipage}
\caption{G123.07-6.31 (a)HCO$^+$(1-0) map grid. (b)Spectral towards (0,0) (left), and (1$\arcmin$,-1$\arcmin$) (right) the dashed line marked the source systemic velocity which was derived from gauss fit of C$^{18}$O(1-0). (c)Position-velocity diagrams of optically thick  HCO$^+$(1-0) line (top) and optically thin C$^{18}$O(1-0) line (bottom), contour levels are 0.4K ($>3\sigma$), 0.7K, 1.0K, 1.3K, 1.7K, 2.1K, 2.5K, 2.9K and 0.3K ($>3\sigma$), 0.5K, 0.7K, 0.8K, 0.9K, respectively.}
\end{figure}

\begin{figure}
%%\vspace{2mm}
\centering
\includegraphics[width=8.cm]{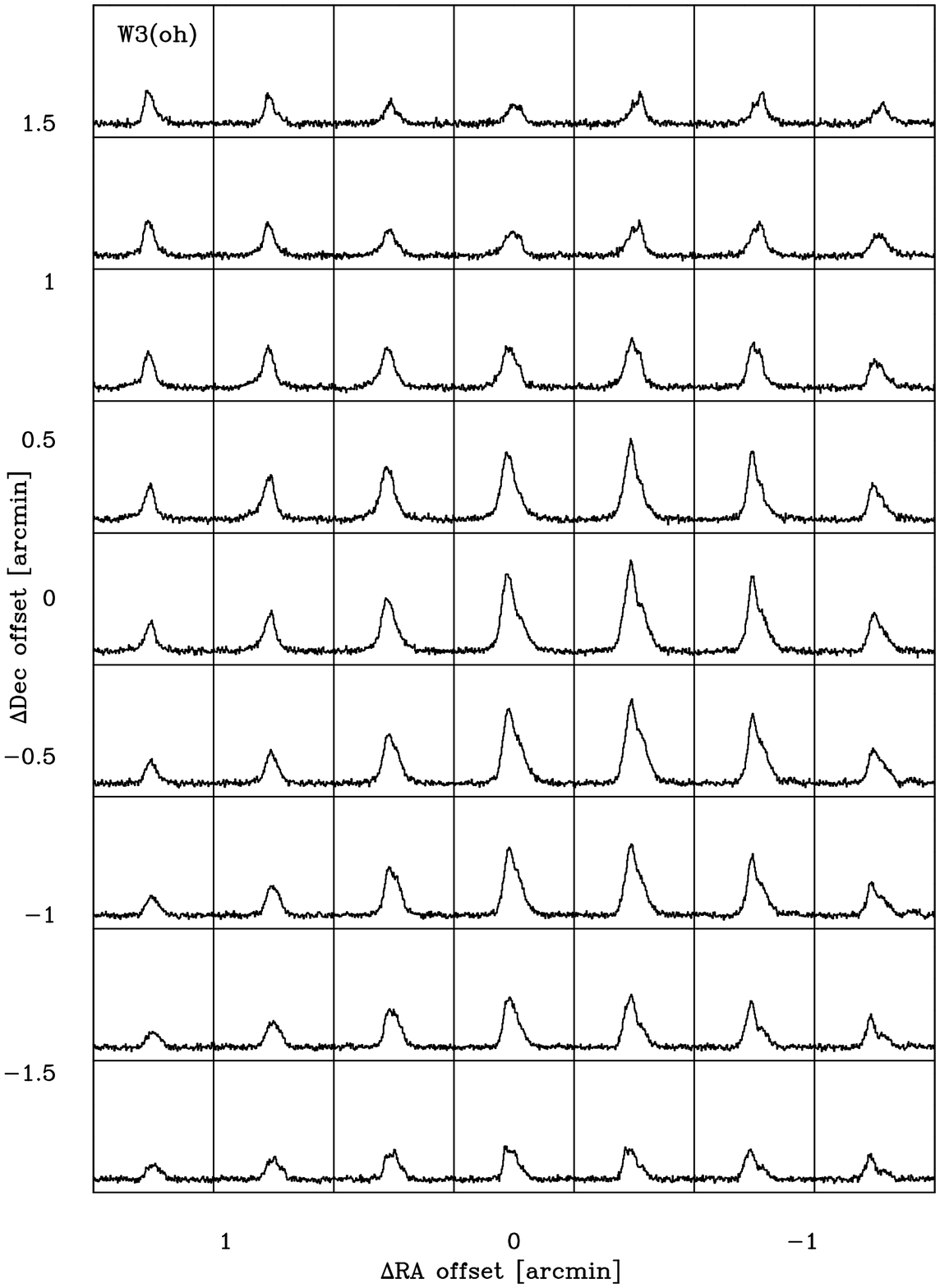}\\
\includegraphics[width=5cm]{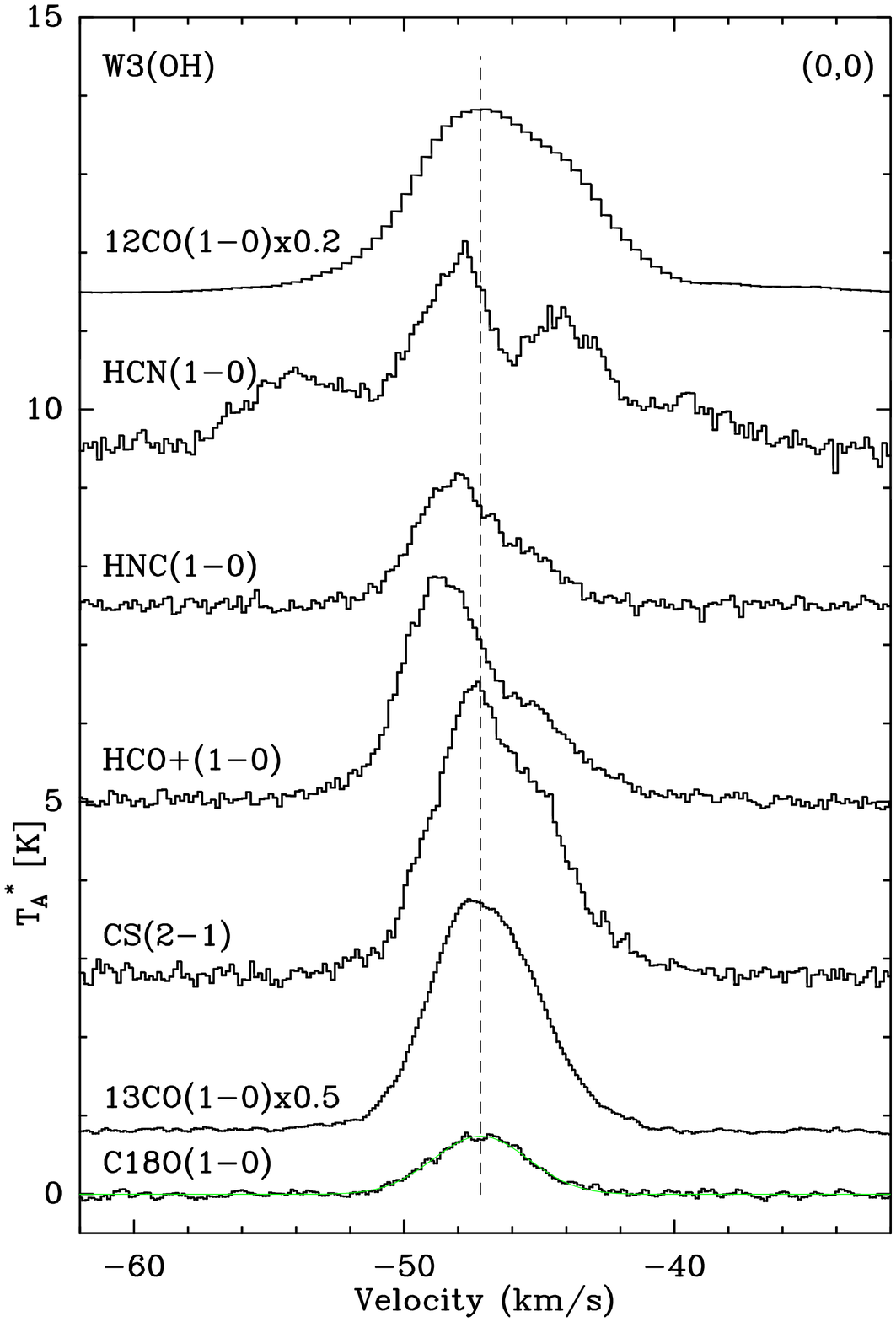}
\caption{W3(OH), the top and bottom panels present the center parts that show obvious 
asymmetric profiles of HCO$^+$(1-0) map grid and all of spectral lines observed  
towards (0,0), respectively. In the top panel, the velocity scale ranges from -62~km~s$^{-1}$ to -32~km~s$^{-1}$ the same as the bottom panel, the temperature scale ranges from -0.5~K to 4.4~K.}
\end{figure}

\begin{figure}
%%\vspace{2mm}
\centering
\includegraphics[width=8.cm]{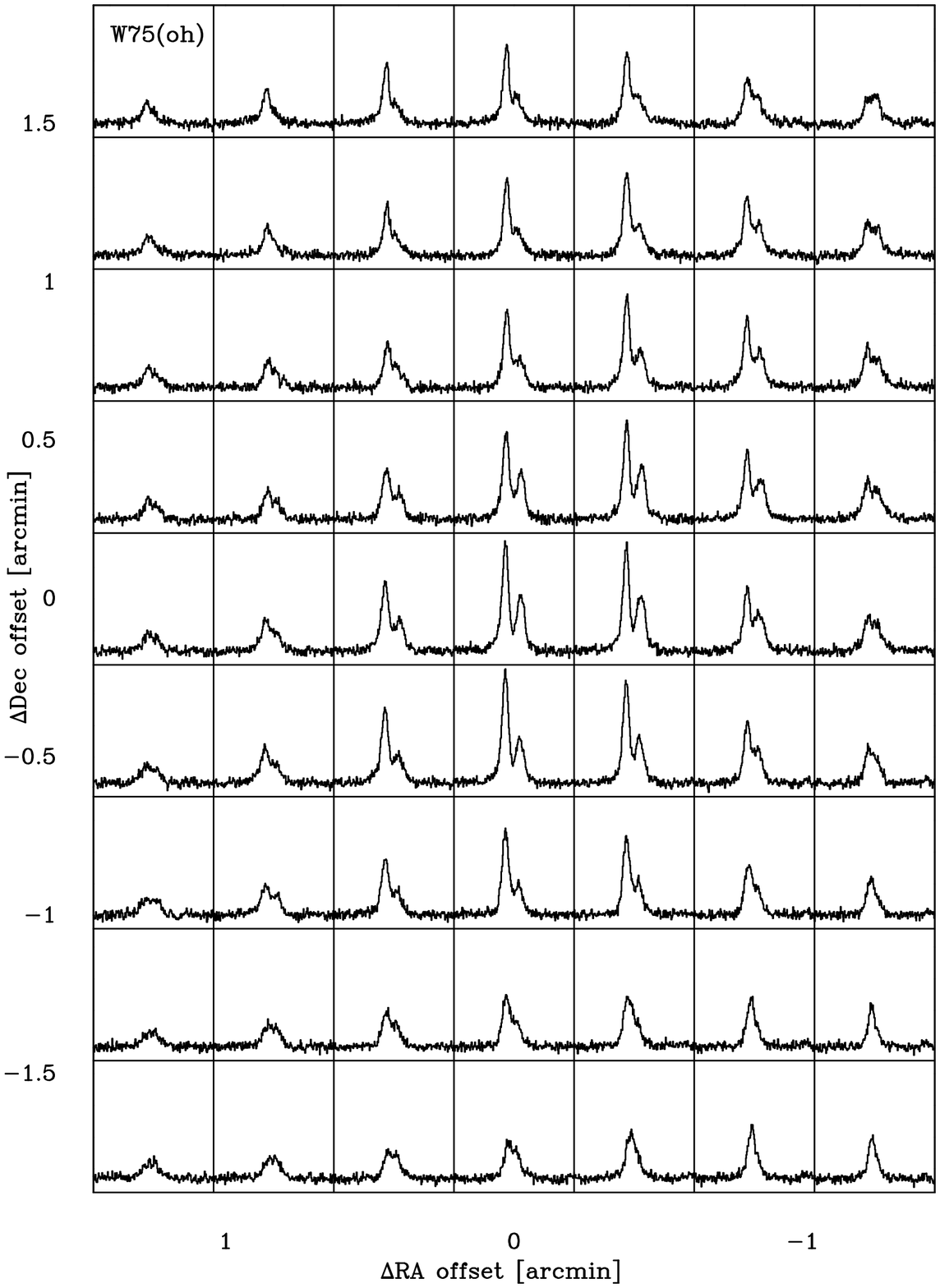}\\
\includegraphics[width=5cm]{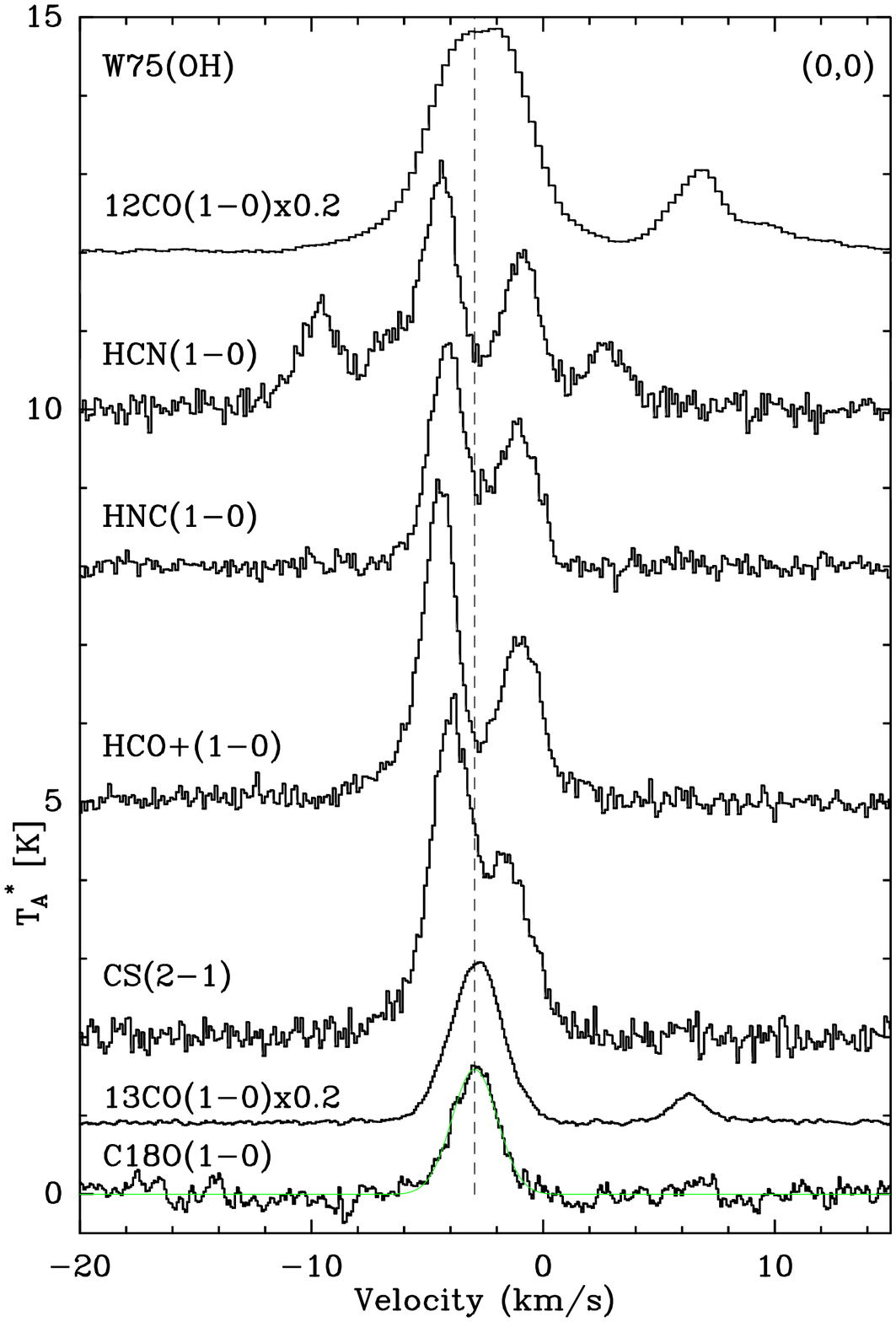}
\caption{W75(OH), the top and bottom panels present the center parts that show obvious 
asymmetric profiles of HCO$^+$(1-0) map grid and all of spectral lines observed  
towards (0,0), respectively. In the top panel, the velocity scale ranges from -20~km~s$^{-1}$ to 15~km~s$^{-1}$ the same as the bottom panel, the temperature scale ranges from -0.5~K to 4.4~K.}
\end{figure}

\begin{figure}
%%\vspace{2mm}
\centering
\includegraphics[width=8cm]{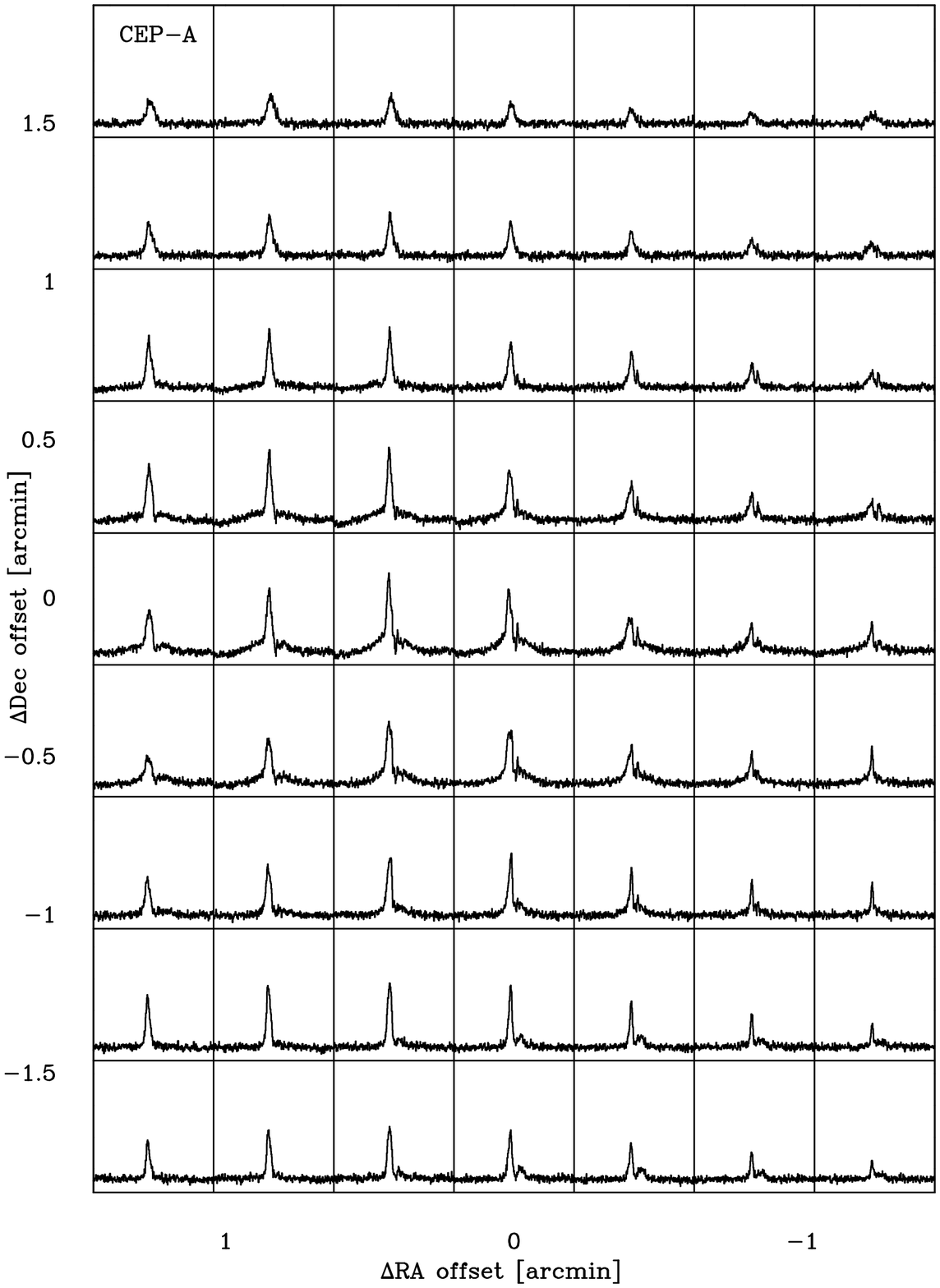}\\
\includegraphics[width=5cm]{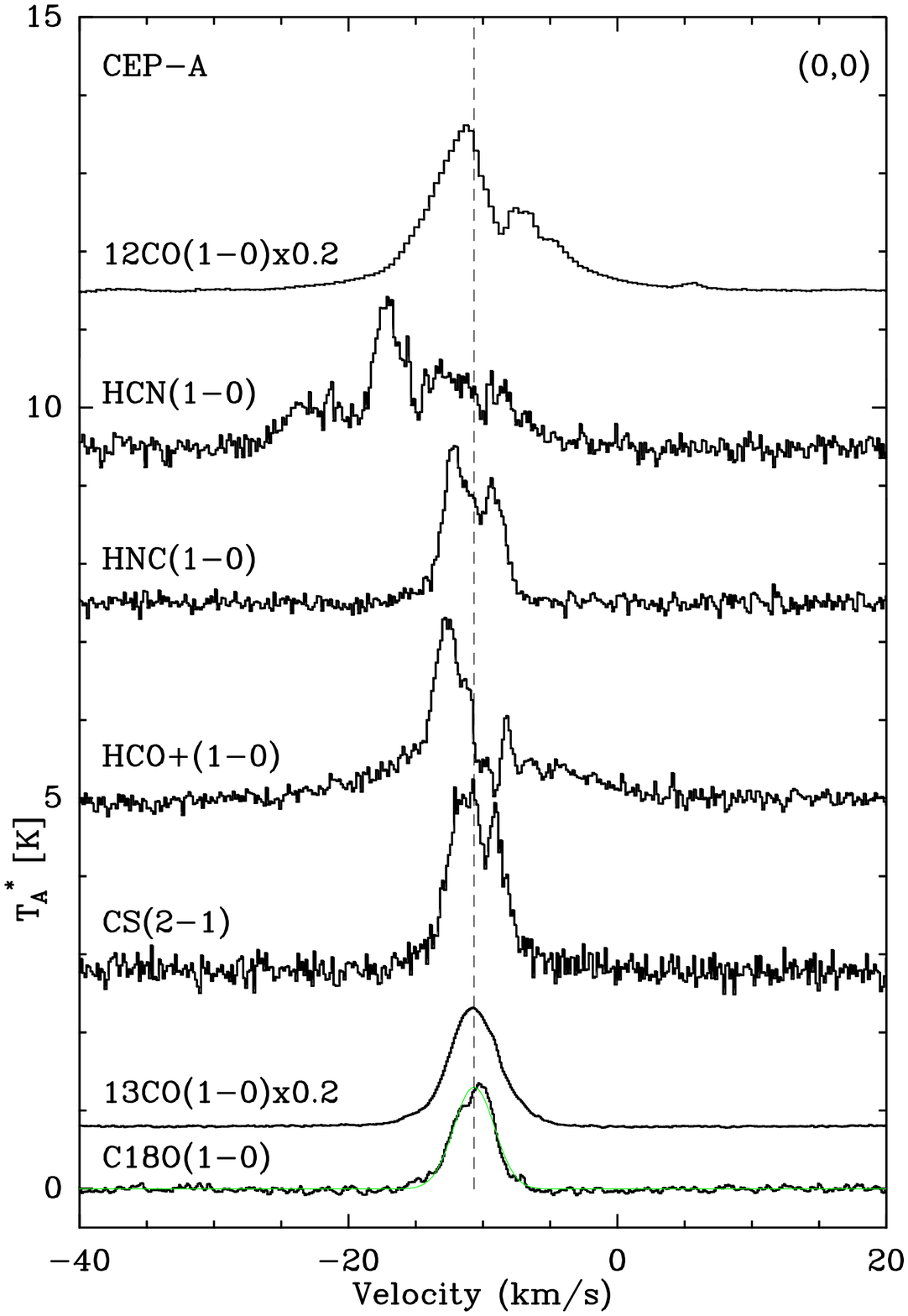}
\caption{CEP-A, the top and bottom panels present the center parts that show obvious 
asymmetric profiles of HCO$^+$(1-0) map grid and all of spectral lines observed  
towards (0,0), respectively. In the top panel, the velocity scale ranges from -40~km~s$^{-1}$ to 20~km~s$^{-1}$ the same as the bottom panel, the temperature scale ranges from -0.5~K to 4.4~K. }
\end{figure}

\begin{figure}
%%\vspace{2mm}
\centering
\includegraphics[width=8.cm]{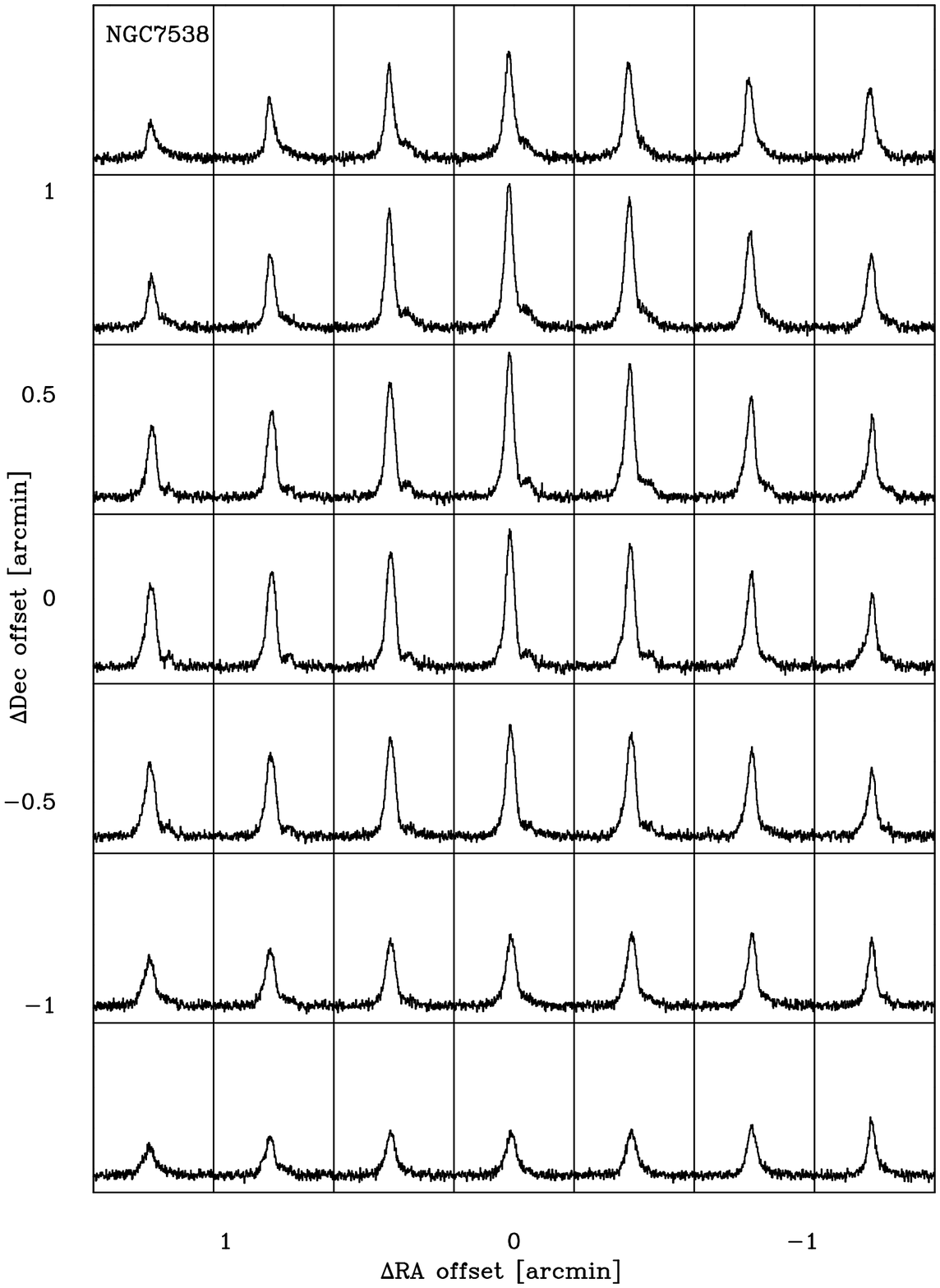}\\
\includegraphics[width=5cm]{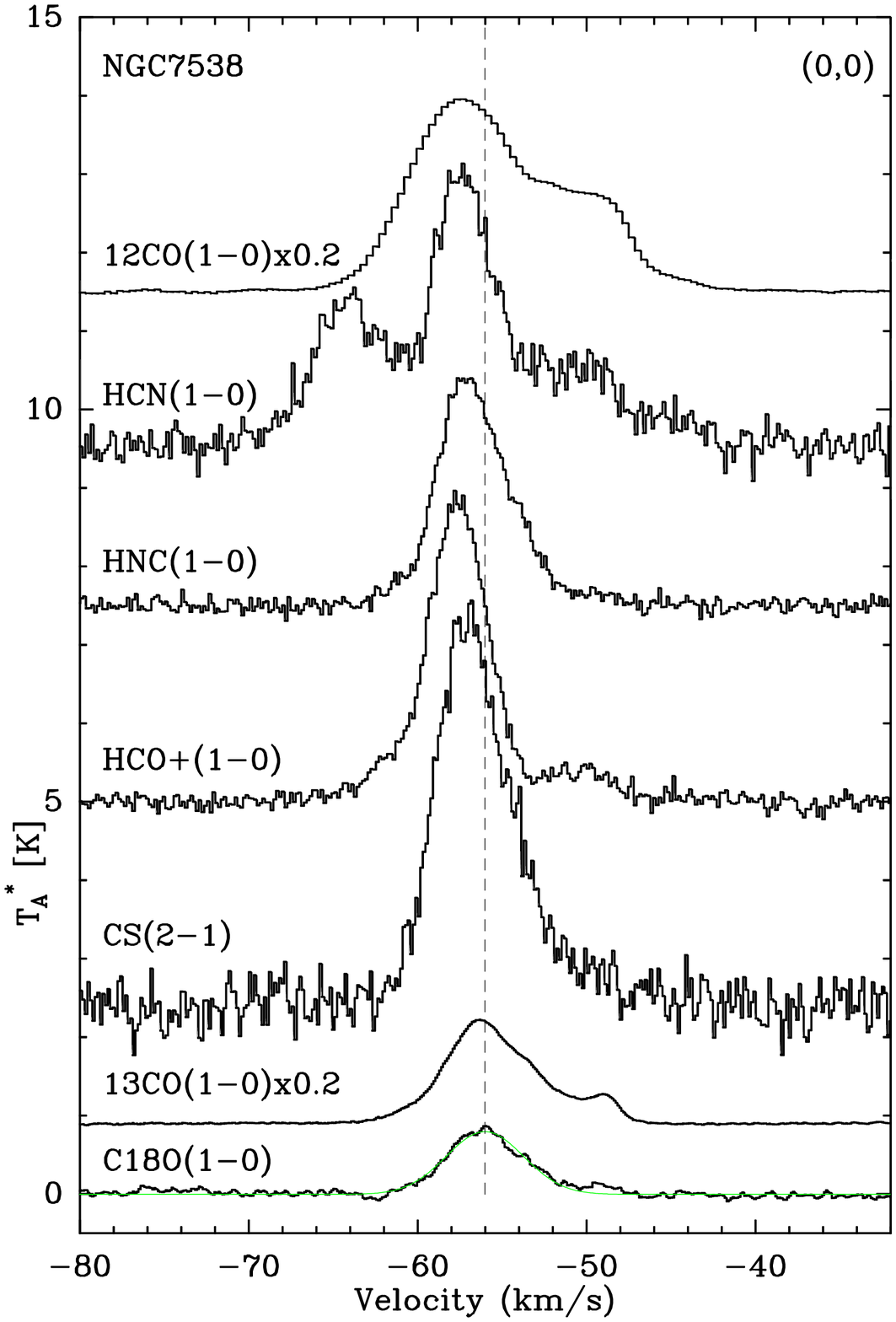}
\caption{NGC7538, the top and bottom panels present the center parts that show obvious 
asymmetric profiles of HCO$^+$(1-0) map grid and all of spectral lines observed  
towards (0,0), respectively.  In the top panel, the velocity scale ranges from -80~km~s$^{-1}$ to -32~km~s$^{-1}$ the same as the bottom panel, the temperature scale ranges from -0.5~K to 4.4~K.}
\end{figure}

\begin{figure}
%%\vspace{2mm}
\centering
\includegraphics[width=8.cm]{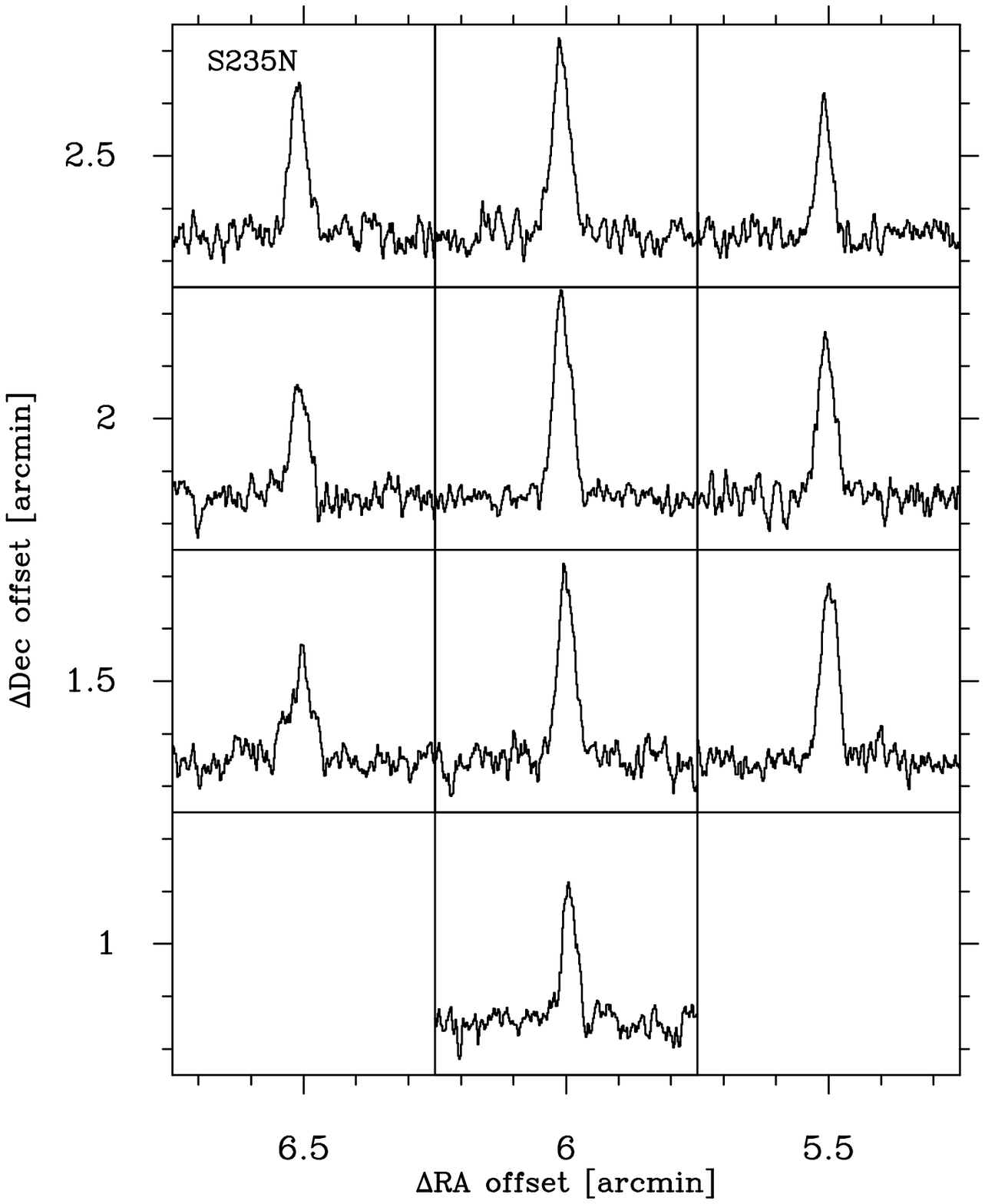}\\
\includegraphics[width=5cm]{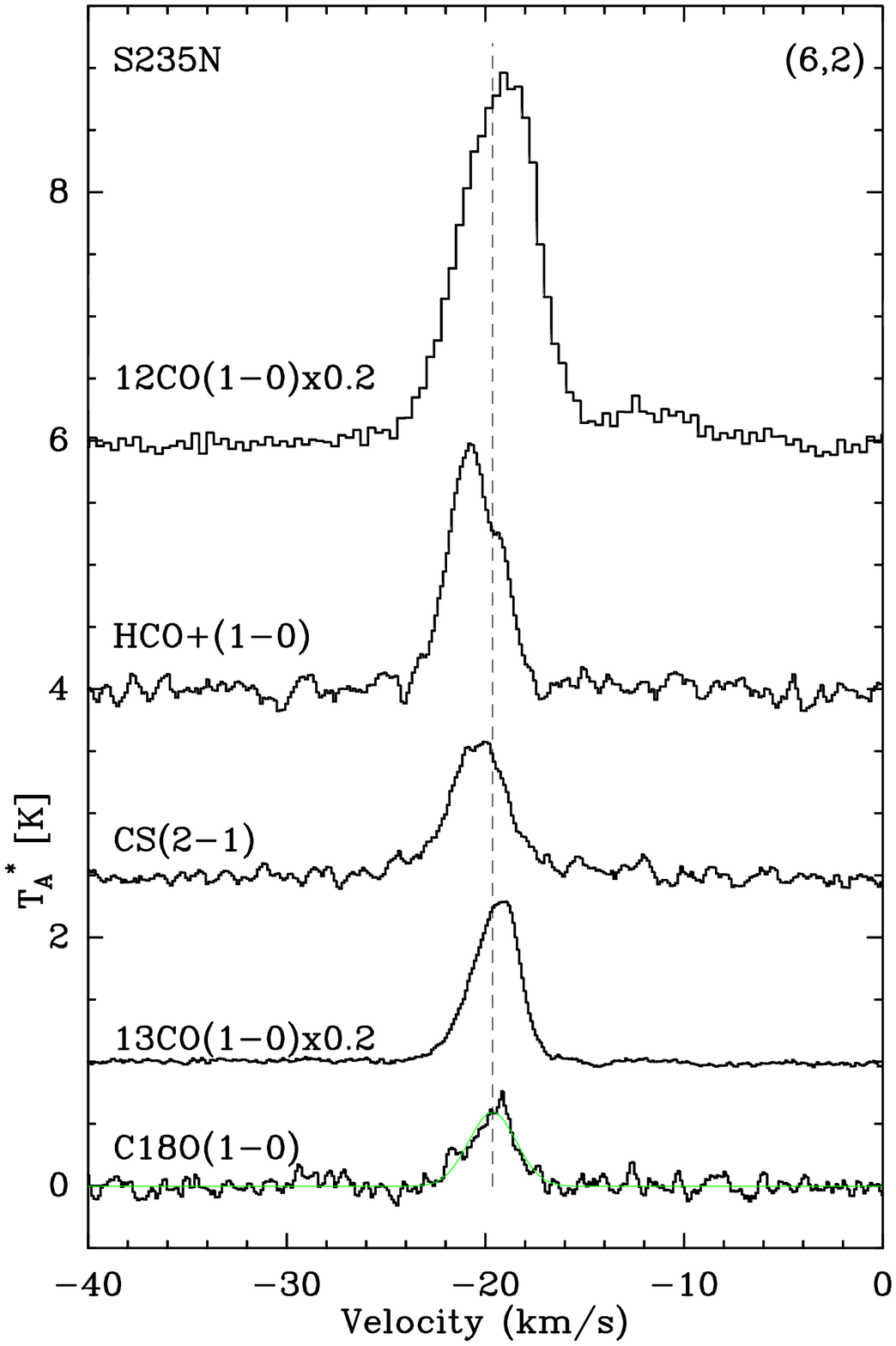}
\caption{S235N, the top and bottom panels present HCO$^+$(1-0) map grid and all of spectral lines observed  
towards (6$\arcmin$,2$\arcmin$), respectively.  In the top panel, the velocity scale ranges from -40~km~s$^{-1}$ to 0~km~s$^{-1}$ the same as the bottom panel, the temperature scale ranges from -0.5~K to 2~K.}
\end{figure}

 A noticeable feature in Figure~9 is that the HCO$^+$(1-0) lines generally have higher
intensity ratios T$_{\rm B}/T_{\rm R}$ than that in CS(2-1) for the six infall
candidates; typically$\sim$2.2 for HCO$^+$(1-0) and $\sim$1.4 for
CS(2-1). The difference is significantly larger than any possible
errors produced in determining  T$_{\rm B}/T_{\rm R}$. A simple analytic
model for collapse (Myers et al. 1996) predicted that the ratio of
blue to red component (T$_{\rm B}/T_{\rm R}$)  increases with infall speed
(V$_{in}$), which assumes that the cloud consists of two uniform
paralleled components without rotation and the infall speed of the
gas is less than the velocity dispersion $\sigma$. Thus, it is likely that HCO+ and CS trace different infall speeds
given their different intensity ratios. This also suggests that the infall speeds 
derived from HCO$^+$(1-0) and CS(2-1) might be different. Of course, we do not expect the derived infall speeds to be same because
 different infall tracers might trace different spatial components of the clouds along the line of sight. From the mapping observations, it is obvious that
the infall speed is also different in different spatial components perpendicular 
to the line of sight. Thus, these two lines do not necessarily provide duplicate information. Rotation might cause blue asymmetric line profile at one side 
of the rotation axis,  but at the same time the red asymmetry 
could appear at the other side of the rotation axis. Based 
on this argument, we could exclude the rotation as the main contributor. We did not do any model fit towards these sources in this paper, 
some modelling efforts will need to be developed in order to derive the infall motion and mass
infall rate in future work.

\begin{figure}
%%\vspace{2mm}
\centering
\includegraphics[scale=0.5]{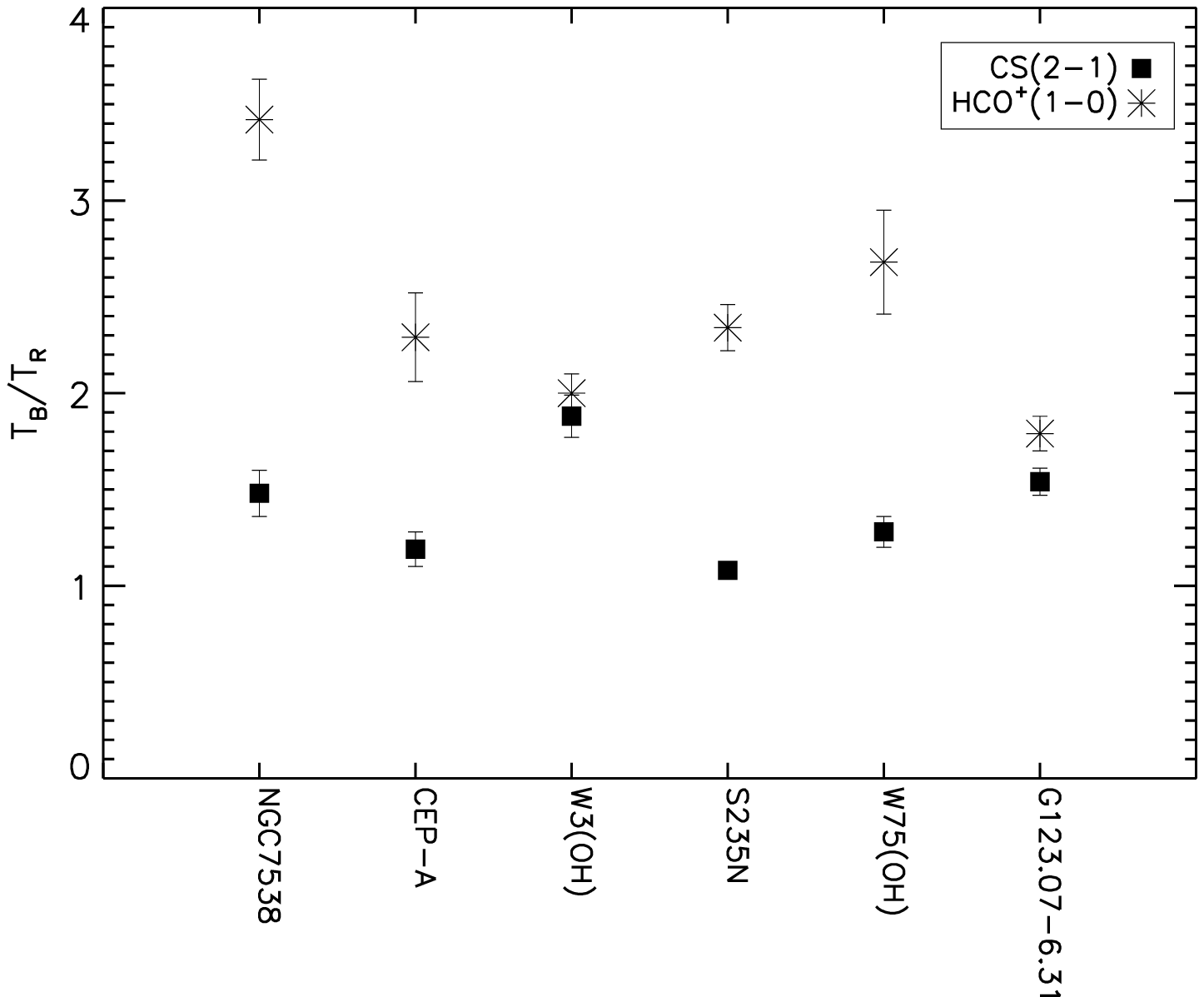}
\caption{Intensity ratios T$_{\rm B}/T_{\rm R}$ in HCO$^+$(1-0) and CS(2-1) for the six infall candidates.}
\end{figure}

\section{Discussion}

 As presented in the previous section, an obvious asymmetry parameter difference was shown between HCO$^+$(1-0) and CS(2-1) in Figs.~2 and 9. HCO$^+$(1-0) line seems to be a better tracer for the study of infall motion compared with CS(2-1). In general, the characteristic blue profile will appear only if the molecular line tracer has a suitable optical depth and critical density. Both HCO$^+$(1-0) and CS(2-1) are optically thick lines, but HCO$^+$(1-0) is more opaque than CS(2-1) (Evans 2003). The critical density of the two lines is also different. At T$_k$=100$\:$K, the critical density of CS(2-1) is $3.9\times10^{5}cm^{-3}$ (Evans 1999), while the critical density of HCO$^+$(1-0) is $1.9\times10^{5} cm^{-3}$ (Evans 1999; cf. $5.0\times10^{4} cm^{-3}$ Fuller et al. 2005). A natural explanation for the asymmetry parameter difference between CS(2-1) and HCO$^+$(1-0) is that infall motions are more easily detected in the lower density outer region of the dense core. In the denser regions, the infall motions might be significantly decelerated by thermal pressures, outflows, and other feedbacks from the massive young objects. 

Nevertheless, HCN(3-2) has a much higher critical density ($6.8\times10^7 cm^{-3}$ at T$_k=100\:K$) than that of CS(2-1), yet it still shows a high inflow motion detection rate (Wu \& Evans 2003). In the fifteen sources that overlap between our survey and that of Wu \& Evans (2003), the profiles derived from HCO$^+$(1-0) are more consistent with that derived from HCN(3-2) than that of CS(2-1) (see Table~3.). The difference between the optical depth of the two molecular lines might be the more important reason than the difference in the critical density of the two lines. A recent model about blue asymmetry in low-mass star-forming region suggests that the line asymmetry is strongly dependent on the abundance distribution (Tsamis et al. 2008), which is proportional to the optical depth.

In the fifteen overlap sources, more blue and red profiles were presented by Wu \& Evans (2003) than that of ours. Their HCN(3-2) observations were carried out in a single pointing toward dense cores using the CSO 10.4$\:$m telescope with resolution about 28$\arcsec$. The higher spatial resolution enables them to better resolve the inner small regions of inward motions. In particular, G19.61-0.23 shows blue profile in both HCO$^+$(1-0) of our results and HCO$^+$(4-3) of the JCMT observations with resolution about 15$\arcsec$ (Klaassen \& Wilson. 2007), but no obvious asymmetry in HCN(3-2) (Wu \& Evans 2003). S88 shows blue profile in HCO$^+$(1-0) over a 2$\arcmin\times2\arcmin$ mapping region in our observations whereas it shows red profile in a single pointing HCN(3-2) observation of Wu \& Evans (2003), yet it shows no obvious asymmetry in single pointing HCO$^+$(4-3) observation of Klaassen \& Wilson (2007). 
 In a recent survey, Fuller et al. (2005) also found that the detection rate of infall motion decreases with increasing source distance. In short, the source distance, half power beam width (HPBW) of the molecular transition line probes, and the size scale of infall region could all affect the infall detection rate.

  Even though our observations (resolution $\sim1\arcmin$) did not have the resolving power to probe a linear scale $<$0.16$\:$pc, the mapping observation of multiple optically thick molecular line could still be useful to give evidence to large scale infall signatures in dense cores. Six infall candidates were identified in our mapping observations, in which large scale ($>$0.2$\:$pc) infall signatures were detected. Among them, NGC7538 has been previously well studied by Sandell et al. (2005). Their high resolution BIMA HCO$^+$(1-0) observation suggests that an embedded protostar embedded in the core is still in a phase of active accretion. All these might suggest that large scale infall motions are closely related to the kinematics of the inner materials of the cores. A large scale infall might play a certain role in providing some materials for the innermost accretion.

Actually, outflow makes an important contribution to the line wing emissions.
But high quality spectra of much higher signal-to-noise ratios are needed to make 
better judgement in line wings.  Following previous similar studies and our observational results, 
we here believe that the line asymmetry is mainly contributed by infall motion. And to a certain extent, large scale infall signature at least provides an indirect evidence for the inner accretion. Protostellar accretion could be the main physical process responsible for massive star formation. Some models suggest that the accretion must halt before a visible UC H{\sc ii} phase (eg. Garay \& Lizano 1999), while other models suggest that an ionized accretion can continue through an H{\sc ii} region (eg. Keto 2003). Four out of our six infall candidates have already formed UC H{\sc ii} region (W3(OH), CEP-A, NGC7538) and H{\sc ii} region (G123.07-6.31), yet they still show large scale infall signatures. Although we don't know whether infall motions halt at smaller scale of the cores, these examples at least at large scale appeared to be consistent with the ionized accretion theory of Keto (2003).     
    
    Comparing with other surveys, our observations show only two (8\%) red asymmetry detection in CS(2-1) and no red asymmetry detection in HCO$^+$(1-0) (vs. 21\% in HCN(3-2) observation of Wu \& Evans (2003); 16\% in HCO$^+$(1-0) observation of Fuller et al. (2005); 10\% in HCO$^+$(1-0) of Wu et al. (2008)). The likely reason is beam dilution and/or the fact that the samples of all surveys  until now are still rather limited (typically $\sim$30, e.g., 28 sources of Wu \& Evans (2003); 23 sources of Klaassen \& Wilson (2007), vs. 77 sources of Fuller et al. (2005)). Of course, these are just some face values and the exact percentage should be affected by resolution. But it is rather arbitrary to scale the resolution in comparing to other observations. S87 is such an interesting source which shows different asymmetry in CS(2-1) and HCO$^+$(1-0) at different offsets (so might be G35.20-0.74). With our telescope resolving power, it's hard to give a reasonable explanation for the different asymmetry in different positions. The possible explanation also include cloud-cloud collision (e.g. S87, Xue \& Wu 2008). Oscillating was also suggested in some low mass cores, e.g., B68 (Redman et al. 2006) and perhaps this might be possible in these massive cores as well. High resolution observations are required towards these sources in order to learn more details about the kinematics of massive star-forming cores. Unlike blue profile, the origin of red profile is still unclear. A simple explanation is that the red profile could be caused by outward motions.

\section{Summary}

We performed a systematic multiple transition 3$\,$mm molecular
line single-pointing and mapping survey towards 29 massive
star-forming cores to search for the signatures of the inward
motions. Up to seven different transitions, optically thick lines
HCO$^+$(1-0), CS(2-1), HNC(1-0), HCN(1-0), $^{12}$CO(1-0) and
optically thin lines C$^{18}$O(1-0), $^{13}$CO(1-0) were observed
towards each source. The major results found from these
observations include the following:

\begin{enumerate}

\item The normalized velocity differences ($\delta$V$_{\rm CS}$,
$\delta$V$_{\rm HCO^{+}}$) between the peak velocities of
optically thick lines and optically thin C$^{18}$O(1-0) for each source
were derived. We found a
significant difference in the incidence of blue-shifted line
asymmetry between CS(2-1) and HCO$^+$(1-0). The HCO$^+$(1-0) line
shows the highest occurrence of obvious asymmetric feature. The
optical depth difference between the two molecular lines may be
responsible for the anomaly. HCO$^+$(1-0) line appears to be one
of the best inward motion tracers in massive cores whereas CS(2-1)
is not a sensitive tracer to signify the infall motion.

\item Blue profile dominance appeared in both CS(2-1) and
HCO$^+$(1-0) in most of our sample. This implies the predominance
of the inward motion in the massive star-forming cores, which is
similar to that in low mass star forming regions. All those may
suggest that protostellar accretion could be the main physical
process responsible for massive star formation.

\item The mapping observations of multiple line transitions enable
us to identify six strong infall candidates (G123.07-6.31,
W75(OH), S235N, CEP-A, W3(OH), NGC7538). Signatures of extended
inward motions are presented in these massive star-forming cores
with possibly strong infall.  The infall signature is extended up
to a linear scale  $>0.2 \:$pc.

\end{enumerate}

\section*{Acknowledgments}

We are grateful to all staff of the 14$\;$m telescope of Purple
Mountain Observatory  for their dedicated assistance, H.J. Ma for the use of her FCRAO data. We
thank Profs. J.Z. Wang, X.Z. Zheng, \& Y.F. Wu, Dr. J.W. Wu, and Mr. M. Fang for useful
discussions. We also thank the anonymous referee for a very helpful report and comments that help improved the paper. 
Research for this project is supported by NSFC Distinguished Young Scholars (\#10425313) and Chinese Academy of Sciences' Hundred Talent Program.

\appendix
\section[]{IRAC and CO line comparison in S235}

The recently available high-resolution and very sensitive mid-infrared and far-infrared images of the Spitzer telescope 
can provide a new insight in directly locating the distribution of dust obscured massive stars in the dense cores.    
As an example for some well studied dense core sources, we here
compare our molecular line observations with the Spitzer IRAC
images in S235. The Spitzer IRAC data were retrieved from Spitzer Science Center\footnote{
http://ssc.spitzer.caltech.edu}. IRAC 4 bands data of S235 (PI G. Fazio, PID$=$201) covering$\sim$0.25 
deg$^2$ reveal a complex star-forming region.

The S235 star-forming region is a well known H{\sc
ii} region that has been widely studied. It has been identified as
a water maser (Lo, Burke, \& Haschick 1975) with infrared (e.g.
Krassner, Pipher \& Sharpless 1979; Felli et al. 1997; Allen et
al. 2005), radio continuum (Felli et al. 1997, 2006), dust
continuum emission (Lee, Young \& Shirley 2002), and many
molecular line emission (e.g. Evans et al. 1981; Felli et al.
1997; Lee, Young, \& Shirley 2002).

Our CO maps show that three $^{13}$CO(1-0) cores are well matched
with three infrared emission regions in Figure~A1. We observed
several points in these three cores in CS(2-1) and HCO$^+$(1-0),
only HCO$^+$(1-0) show obvious blue profile around position (6\arcmin,2\arcmin)
marked dash open circle.  Class I and class II sources in S235 were
identified by Allen et al. 2005, and their results show that
few YSOs are distributed in the North-West extended core  and only two
class I sources are distributed around position (6\arcmin,2\arcmin), whereas many more YSOs
are distributed in the other two cores. All these may imply that
the region around position (6\arcmin,2\arcmin) is still on the very early stage
of star formation.
\begin{figure}
%%\vspace{2mm}
\centering
\begin{minipage}[c]{0.5\textwidth}
\centering
\includegraphics[scale=0.5]{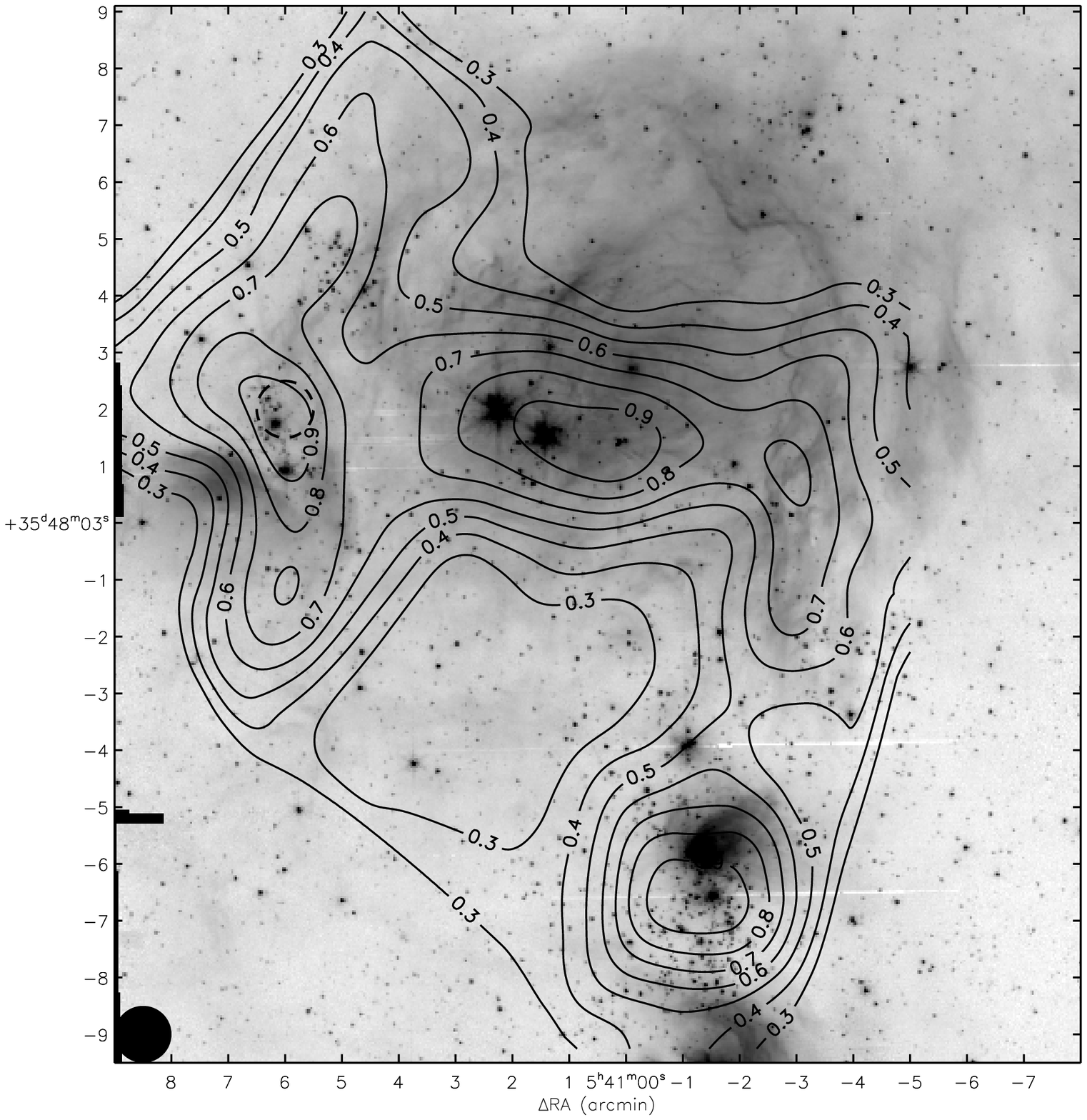}
\end{minipage}
\caption{The $^{13}$CO(1-0) contour map overlaid on the Spitzer
IRAC 4.5$\:\mu$m mosaic of S235. Contour levels start at 30\% of
the peak $^{13}$CO(1-0) integrated intensity and increase by 10\%
at each interval.}
\end{figure}

\label{lastpage}


\begin{thebibliography}{99}
\bibitem[\protect\citeauthoryear{Allen et al.}{2005}]{al} Allen L.E., Hora J.L., Megeath S.T., Deutsch L.K., Fazio G.G., Chavarria L., Dell R.W. 2005, IAUS, 227, 352
\bibitem[\protect\citeauthoryear{Beuther et al.}{2002}]{be} Beuther H., Schilke P., Menten K.M., Motte F., Sridharan T.K., Wyrowski F., 2002, ApJ, 566, 945
\bibitem[\protect\citeauthoryear{Beltr\'an et al.}{2006}]{bel} Beltr\'an M.T., Cesaroni R., Codella C., Testi L., Furuya R.S., Olmi L., 2006, Nat, 443, 427
\bibitem[\protect\citeauthoryear{Bonnell et al.}{1998}]{bo} Bonnell I.A., Bate M.R., Zinnecker H., 1998, MNRAS, 298, 93
\bibitem[\protect\citeauthoryear{Bonnell \& Bate}{2006}]{bob} Bonnell I.A., Bate M.R., 2006, MNRAS, 370, 488
\bibitem[\protect\citeauthoryear{Carral et al.}{1999}]{ca} Carral P., Kurtz S., Rodr¨ªguez L.F., Mart J., Lizano S., Osorio M., 1999, RMxAA, 35, 97
%\bibitem[\protect\citeauthoryear{Codella et al.}{2006}]{co} Codella C., Viti S., Williams D.A., Bachiller R., 2006, ApJ, 644, L41
\bibitem[\protect\citeauthoryear{Dickel \& Auer}{1994}]{di} Dickel H.R., Auer L.H., 1994, ApJ, 437, 222
\bibitem[\protect\citeauthoryear{Eiroa, Elasser \& Lahulla}{1979}]{ei} Eiroa C., Elasser H., Lahulla J.F., 1979, A\&A, 74, 89
\bibitem[\protect\citeauthoryear{Evans}{1999}]{evan} Evans N.J., 1999, ARAA, 37, 311
\bibitem[\protect\citeauthoryear{Evans}{2003}]{eva} Evans N.J., 2003, in Chemistry as a Diagnostic of Star Formation, eds. C.L. Curry \& M. Fich, NRC Press, Ottawa, Canada, 2003, p. 157.
\bibitem[\protect\citeauthoryear{Evans et al.}{1981}]{ev} Evans N.J., II, Beichman C., Gatley I., Harvey P., Nadeau D., Sellgren K., 1981, ApJ, 246, 409
\bibitem[\protect\citeauthoryear{Fazio et al.}{2004}]{fa} Fazio, G.G. et al., 2004, ApJS, 154, 10
\bibitem[\protect\citeauthoryear{Felli et al.}{1997}]{ft} Felli M., Testi L., Valdettaro R., Wang J.-J., 1997, A\&A, 320, 594
\bibitem[\protect\citeauthoryear{Felli et al.}{2006}]{fm} Felli M., Massi F., Robberto M., Cesaroni R., 2006, A\&A, 453, 911
\bibitem[\protect\citeauthoryear{Fuller, Williams \& Sridharan}{2005}]{fu} Fuller G.A., Williams S.J., Sridharan T.K., 2005, A\&A, 442, 949
\bibitem[\protect\citeauthoryear{Garay \& Lizano}{1999}]{ga} Garay G., Lizano S., 1999, PASP, 111, 1049
\bibitem[\protect\citeauthoryear{Gregersen et al.}{2000}]{gr} Gregersen E.M., Evans N.J., II, Mardones D., Myers P.C., 2000, ApJ, 533, 440
\bibitem[\protect\citeauthoryear{Jiang et al.}{2005}]{ji} Jiang Z.B., Tamura M., Fukagawa M., 2005, Nat, 437, 112
\bibitem[\protect\citeauthoryear{Keto}{2003}]{ke} Keto E., 2003, ApJ, 599, 1196
%\bibitem[\protect\citeauthoryear{Keto \& Wood}{2006}]{kw} Keto E., Wood K., 2006, APJ, 637, 850
\bibitem[\protect\citeauthoryear{Klaassen \& Wilson}{2007}]{kl} Klaassen P.D., WiLson C.D., 2007, ApJ, 663, 1092
\bibitem[\protect\citeauthoryear{Krassner, Pipher \& Sharpless}{Krassner et al.}{1979}]{kr} Krassner J., Pipher J.L., Sharpless S., 1979, A\&A, 77, 302
\bibitem[\protect\citeauthoryear{Krumholz \& Bonnell}{2007}]{kb} Krumholz M.R., Bonnell I.A., 2007, astro-ph, 0712, 0828
\bibitem[\protect\citeauthoryear{Lee, Myers \& Tafalla}{Lee et al.}{1999}]{lm} Lee C.W., Myers P.C., Tafalla M., 1999, ApJ, 526, 788
\bibitem[\protect\citeauthoryear{Lee, Myers \& Tafalla}{Lee et al.}{2001}]{lmt} Lee C.W., Myers P.C., Tafalla M., 2001, ApJS, 136, 703
\bibitem[\protect\citeauthoryear{Lee, Myers \& Plume}{Lee et al.}{2004}]{lmp} Lee C.W., Myers P.C., Plume R., 2004, ApJS, 153, 523
\bibitem[\protect\citeauthoryear{Lee, Young \& Shirley}{2002}]{lys} Lee J.-E., Young C.H., Shirley Y.L., Mueller K.E., Evans N.J., II, 2002, ASPC, 267, 377
\bibitem[\protect\citeauthoryear{Lo, Burke \& Haschick}{Lo et al.}{1975}]{lb} Lo K.Y., Burke B.F., Haschick A.D., 1975, ApJ, 202, 81
\bibitem[\protect\citeauthoryear{Ma \& Gao}{2008}]{mg} Ma H.J., Gao Y., Wu J.W., 2008, in prep.
\bibitem[\protect\citeauthoryear{Mardones \& Myers}{1997}]{ma} Mardones D., Myers P.C., 1997, ApJ, 489, 719
\bibitem[\protect\citeauthoryear{Moscadelli \& Cesaroni}{2005}]{mo} Moscadelli L., Cesaroni R., 2005, A\&A, 438, 889
\bibitem[\protect\citeauthoryear{Myers \& Mardones}{1996}]{my} Myers P.C., Mardones D., 1996, ApJ, 465, 133
\bibitem[\protect\citeauthoryear{Redman, Keto \& Rawlings}{Redman et al.}{1996}]{re} Redman M.P., Keto E., Rawlings J.M.C., 2006, MNRAS, 370, 1
\bibitem[\protect\citeauthoryear{Ridge et al.}{2003}]{ri} Ridge N.A., Wilson T.L., Megeath S.T., Allen L.E., Myers P.C., 2003, AJ, 126, 286
\bibitem[\protect\citeauthoryear{Rieke et al.}{2004}]{rieke} Rieke, G.H. et al., 2004, ApJS, 154, 25
\bibitem[\protect\citeauthoryear{Sandell et al.}{2005}]{sa} Sandell G., Goss W.M., Wright M., 2005, ApJ, 621, 839
\bibitem[\protect\citeauthoryear{Shirley, Evans \& Young}{2003}]{sh} Shirley Y.L., Evans N.J., II, Young K.E., Knez C., Jaffe D.T., 2003, ApJ, 149, 375
\bibitem[\protect\citeauthoryear{Shu}{1977}]{sh} Shu F.H., 1977, ApJ, 214, 488
\bibitem[\protect\citeauthoryear{Shu et al.}{1987}]{sa} Shu F.H., Adams F.C., Lizano S., 1987, ARA\&A, 25, 23
\bibitem[\protect\citeauthoryear{Sohn et al.}{2007}]{sl} Sohn J., Lee C. W., Park Y.-S., Lee H.M., Myers P.C., Lee Y., 2007, ApJ, 664, 928
\bibitem[\protect\citeauthoryear{Tofani et al.}{1995}]{to} Tofani G., Felli M., Taylor G.B., Hunter T.R., 1995, A\&AS, 112, 299
\bibitem[\protect\citeauthoryear{Tsamis et al.}{2008}]{ts} Tsamis Y.G., Rawlings J.M.C., Yates J.A., Viti S., 2008, astro-ph, 0803, 0519
\bibitem[\protect\citeauthoryear{Welch et al.}{1988}]{wd} Welch W.J., Dreher J.W., Jackson J.M., 1988, Sci, 238, 1550
\bibitem[\protect\citeauthoryear{Whitney}{2005}]{wh} Whitney B.A., 2005, Nat, 437, 37
\bibitem[\protect\citeauthoryear{Wolf-Chase \& Gregersen}{1997}]{wg} Wolf-Chase G.A., Gregersen E., 1997, ApJ, 479, L67
\bibitem[\protect\citeauthoryear{Wu \& Evans}{2003}]{we} Wu J.W., Evans N.J., II, 2003, ApJ, 592, L79
\bibitem[\protect\citeauthoryear{Wu et al.}{2007}]{wh} Wu Y.F., Henkel C., Xue R., Guan X., Miller M., 2007, ApJ, 669, L37
\bibitem[\protect\citeauthoryear{Wu, Zhu \& Wei}{Zhu et al.}{2005}]{wz} Wu Y.F., Zhu M., Wei Y., 2005, ApJ, 628, L57
\bibitem[\protect\citeauthoryear{Xu et al.}{2006}]{xu1} Xu Y., Reid M.J., Zheng X.W., Menten K.M., 2006, Sci, 311, 54
\bibitem[\protect\citeauthoryear{Xu et al.}{2007}]{xu2} Xu Y., Reid M.J., Menten K.M., Brunthaler A., Zheng X.W., Moscadelli L., 2007, IAUS, 242, 374
\bibitem[\protect\citeauthoryear{Xue \& Wu}{2008}]{xue} Xue R., Wu Y.F., 2008, ApJ, 680, 446
\bibitem[\protect\citeauthoryear{Zhang, Ho \& Ohashi}{Zhang et al.}{1998}]{zh} Zhang Q., Ho P.T.P., Ohashi N., 1998, ApJ, 494, 636
\bibitem[\protect\citeauthoryear{Zhang, Hunter \& Brand}{Zhang et al.}{2001}]{zh} Zhang Q., Hunter T.R., Brand J., 2001, ApJ, 552, L167
\bibitem[\protect\citeauthoryear{Zhou et al.}{1993}]{ze} Zhou S., Evans N.J., II, Koempe C., Walmsley C.M., 1993, ApJ, 404, 232
\bibitem[\protect\citeauthoryear{Zhou et al.}{1994}]{zek} Zhou S., Evans N.J., II, Koempe C., Walmsley C.M., 1994, ApJ, 421, 854
\end{thebibliography}
\end{document}